\documentclass[twocolumn,floatfix,pra,aps,showpacs,superscriptaddress]{revtex4-1}
\usepackage{graphicx,graphics}
\usepackage{amsmath}
\usepackage{amssymb}
\usepackage{epstopdf}
\usepackage{amstext}
\usepackage{amsthm}
\usepackage{amsfonts}
\usepackage{latexsym}
\usepackage{array}
\usepackage{color}

\begin{document}

\title{Multidimensional Josephson vortices in spin-orbit coupled Bose-Einstein 
condensates: snake instability and decay through vortex dipoles} 

\author{A. Gallem\'{\i}}
\affiliation{Departament d'Estructura i Constituents de la Mat\`{e}ria, Facultat de F\'{\i}sica, Universitat de Barcelona, E--08028
Barcelona, Spain}
\affiliation{Institut de Nanoci\`encia i Nanotecnologia de la Universitat de 
Barcelona, IN$\,^2$UB, E--08028 Barcelona, Spain}
\author{M. Guilleumas}
\affiliation{Departament d'Estructura i Constituents de la Mat\`{e}ria, Facultat de F\'{\i}sica, Universitat de Barcelona, E--08028 Barcelona, Spain}
\affiliation{Institut de Nanoci\`encia i Nanotecnologia de la Universitat de Barcelona, IN$\,^2$UB, E--08028 Barcelona, Spain}
\author{R. Mayol}
\affiliation{Departament d'Estructura i Constituents de la Mat\`{e}ria, Facultat de F\'{\i}sica, Universitat de Barcelona, E--08028 Barcelona, Spain}
\affiliation{Institut de Nanoci\`encia i Nanotecnologia de la Universitat de Barcelona, IN$\,^2$UB, E--08028 Barcelona, Spain}
\author{A. Mu\~noz Mateo}
\affiliation{Departament d'Estructura i Constituents de la Mat\`{e}ria, Facultat 
de F\'{\i}sica, Universitat de Barcelona, E--08028
Barcelona, Spain}

\date{\today}

\begin{abstract}
We analyze the dynamics of Josephson vortex states in two-component 
Bose-Einstein condensates with Rashba-Dresselhaus spin-orbit coupling by 
using the Gross-Pitaevskii equation. In 1D, 
both in homogeneous and harmonically trapped systems, we report on stationary 
states containing 
doubly charged, static Josephson vortices. In 
multidimensional systems, we find stable Josephson vortices in a regime of parameters typical 
of current experiments with $^{87}$Rb atoms. In addition, we discuss the 
instability regime of Josephson vortices in disk-shaped condensates, where the 
snake instability operates and vortex dipoles emerge. We study the rich 
dynamics that they exhibit in different regimes of the spin-orbit 
coupled condensate depending on the orientation of the Josephson vortices.

\end{abstract}


\maketitle

\section{Introduction}

In the last years, new experimental techniques developed in the field of 
ultracold atoms have allowed the study of challenging physical phenomena by quantum 
simulation. Among such techniques, the ability to engineer artificial magnetism 
by means of synthetic gauge fields \cite{Dalibard2011,Goldman2014} is one of 
the most promising topics. In particular, 
spin-orbit (SO) coupling can be simulated in Bose-Einstein condensates of 
ultracold gases (BEC) by laser coupling the momentum and the internal 
degrees of freedom (pseudo-spin) of neutral particles 
\cite{Stanescu2008,Lin2009,Lin2011,Galitski2013}. 
Both experimental and theoretical research have explored diverse systems, 
including 2D configurations \cite{Sinha2011}, optical lattices 
\cite{Hamner2015,Zhang2013}, or ring geometries \cite{Chen2014,Mivehvar2015}; 
and new phenomena have been 
analyzed, such as vortex arrays \cite{Radic2011}, half-quantum vortex states 
with Rashba coupling \cite{Ramachandhran2012}, roton instability with
Rashba-Dresselhaus one \cite{Martone2012}, or external Josephson oscillations 
\cite{Zhang2012b,GarciaMarch2014}.

In the first experimental realization of SO coupling in BECs, an equal amount 
of Rashba-Dresselhaus coupling has been implemented in a two-component 
(pseudospin-$1/2$) condensate \cite{Galitski2013}. This system is characterized 
by an inter-component Raman coupling of frequency $\Omega$, and an 
intra-component 
momentum shift $\pm \hbar k_L$. For homogeneous condensates of this kind, it 
has been theoretically demonstrated that such coupling leads to three different 
phases, namely, stripe, plane-wave, and single-minimum phases, whose
appearance depends on the values of the homogeneous density, 
inter and intraspecies interaction and the linear, 
internal coupling between components \cite{Li2012a,Li2012b,Martone2012,Li2015}.

Since the Raman coupling configures an internal, long Josephson junction 
\cite{Williams1999a,Williams1999b,Gross2010}, 
characteristic effects of Josephson dynamics are expected to occur.
In comparison with short Josephson junctions connecting BECs 
\cite{Smerzi1997,Raghavan1999,Albiez2005,Levy2007}, the 
physics of the long Josephson links, focusing on the local properties 
of the system along a uniform junction, provides a richer variety of dynamical 
processes. In these systems, in the absence of spin-orbit coupling ($k_L=0$), 
and for small tunneling frequencies, there exist metastable domain walls in the 
relative phase of the coupled components \cite{Son2002}. Furthermore, 1D 
condensates admit stationary states with non trivial topology in the relative 
phase \cite{Kaurov2005,Kaurov2006}, in analogy with fluxon states in 
superconductors 
\cite{Barone1982}. Such states, referred to as Josephson vortices 
(JVs), are composed of one soliton in each component and have been analyzed in 1D 
homogeneous condensates, both in linear \cite{Kaurov2005,Kaurov2006} and ring geometries 
\cite{Su2013}. In the presence of spin-orbit coupling ($k_L\neq0$), the 
study of solitonic stationary states in 1D systems has found stable JVs in 
homogeneous settings that persist when a harmonic trap is introduced 
\cite{Achilleos2013}. In this latter arrangement, 1D solitonic structures have 
also been dynamically generated by varying the coherent coupling 
\cite{Cao2015}. All of these results have pointed to the existence of 
equivalent topological structures in multidimensional systems that could be 
realized in ultracold gas experiments.

In this work we demonstrate the existence of stable multidimensional JVs, in 
the presence of Rashba-Dresselhaus SO coupling, in harmonically trapped BECs 
with typical parameters of current experiments with ultracold $^{87}$Rb atoms. 
We present JV states, characterized by a deep 
depletion in the total density profile accompanied by a $2\pi$-phase jump 
in the relative phase between components, belonging either to the plane-wave 
or to the single-minimum phases of the SO coupled system. In 1D settings, we 
also report on static doubly charged JVs, for both 
homogeneous and harmonically trapped condensates. Besides, we analyze the decay 
dynamics of  multidimensional JVs, which are unstable when the chemical 
potential is high enough to excite
vortex lines in the system. Our results show that the 
decay scenarios in disk-shaped condensates are different from those of 
dark solitons in scalar elongated condensates where a single solitonic vortex 
\cite{Brand2001,MunozMateo2014} appears at the final stage of the decay. In 
our systems such a role is played by vortex dipoles, one in each condensate 
component, which are linked by domain walls in the relative phase. We show the 
subsequent dynamics of the vortex dipoles due to the interplay between the SO 
coupling and the JV orientation.

In section \ref{sec:System} we introduce our mean-field model for the spinor 
condensate and provide the numerical parameters we have used for our 
simulations. Section \ref{sec:1D} is devoted to analyze 
the features of solitonic states in 1D systems, which we characterize both in 
homogeneous and harmonically trapped condensates in the presence of SO coupling. 
Furthermore, we report on doubly charged Josephson vortex states. In section 
\ref{sec:multiD}, we present stable multidimensional Josephson vortices and 
analyze the decay in the unstable cases through snake instability. The dynamics 
of the emergent vortex dipoles is studied in the different phases of the SO 
coupled condensate. To sum up, we present our conclusions and perspectives for 
future work in section \ref{sec:conclusions}.

\section{System}
\label{sec:System}

We consider two-component condensates in the mean field regime, described by 
the coupled Gross-Pitaevskii equations (GPE) for the condensate wave function 
$\Psi(\mathbf{r},t)=[\Psi_\uparrow(\mathbf{r},t),\Psi_\downarrow(\mathbf{r},
t)]^T$: 

\begin{align}
 i\hbar\left(\frac{\partial}{\partial t}
 -\frac{\hbar k_L}{m}\frac{\partial}{\partial x}\right)\Psi_\uparrow =
 \left(\mathcal{H}_\uparrow+g_{\uparrow\downarrow}|\Psi_\downarrow|^2 \right)
\Psi_\uparrow + \frac{\hbar\Omega}{2}\Psi_\downarrow \nonumber\\
  i\hbar\left(\frac{\partial}{\partial t}
 +\frac{\hbar k_L}{m}\frac{\partial}{\partial x}\right)\Psi_\downarrow =
 \left(\mathcal{H}_\downarrow+g_{\uparrow\downarrow}|\Psi_\uparrow|^2 \right)
\Psi_\downarrow + \frac{\hbar\Omega}{2}\Psi_\uparrow ,
 \label{tdgpe}
\end{align}
where $\mathbf{k}_L=(k_L,0,0)$ and $\Omega$ are the laser wave vector and Raman 
frequency, 
respectively, characterizing the spin-orbit coupling, and 
$\mathcal{H}_i=-\hbar^2/2m \,\vec{\nabla}^2+V_{\rm 
trap}+g_{ii}|\Psi_i|^2$, with $i=\uparrow, 
\downarrow$, is the Gross-Pitaevskii Hamiltonian 
for each single component without density coupling. We will assume that the 
system stays in the miscible regime \cite{Abad2013}, having equal 
intra-component interaction strengths 
$g=g_{\uparrow\uparrow}=g_{\downarrow\downarrow}=4\pi\hbar^2 a/m$, with 
scattering length $a$, and slightly smaller inter-component one 
$g_{\uparrow\downarrow}=4\pi\hbar^2 a_{\uparrow\downarrow}/m \lesssim g$, as it 
occurs in the hyperfine species of $^{87}$Rb atoms.
The confinement of the system will be provided by a 
cylindrically symmetric harmonic trap, $V_{trap}= m ( 
\,\omega_\rho^2 \rho^2+\omega_z^2 z^2)/2$, where $\rho^2=x^2+y^2$, defining an 
aspect ratio $\gamma=\omega_z/\omega_\rho$, and characteristic lengths 
$a_{ho}=a_\rho=\sqrt{\hbar/m\omega_\rho}$ and $a_z=\sqrt{\hbar/m\omega_z}$. 
The stationary states of Eq. 
(\ref{tdgpe}) will be written as $\Psi(\mathbf{r},t)=\exp(-\mu 
t/\hbar)\,[\psi_\uparrow(\mathbf{r}),\psi_\downarrow(\mathbf{r})]^T$, where
$\mu$ is the chemical potential, and the number of particles $N$ will be fixed 
by the wave function normalization $\sum_i\int \psi_i^*(\mathbf{r}) \psi_i 
(\mathbf{r})d\mathbf{r}=N$.

The dimensionless SO number $m\Omega/\hbar k_L^2$, which measures the ratio 
between 
the energy associated to the linear coupling $\hbar\Omega/2$ and the recoil 
energy $E_L=\hbar^2k_L^2/2m$, determines the features of the whole coupling 
between the components of the condensate. In the absence of interactions, the 
ground state of a system with spin-orbit coupling is degenerate for the nonzero 
momenta $\pm \hbar k_L$, given that the referred coupling parameter fulfills 
$m\Omega/\hbar k_L^2<2$; otherwise, there is a single ground state with zero 
momentum. In repulsively interacting condensates one can observe a similar 
trend in the dynamical regimes of the system: the lowest energy state has zero 
momentum, within the single-minimum (SM) phase, for couplings above 
$m\Omega_c/\hbar k_L^2=2\,(1-(g-g_{\uparrow\downarrow})n_T/E_L)$ (where $n_T$ 
is the total density) and non-zero below such 
value \cite{Li2015}. However the physics of the nonlinear system is richer, and
different dynamical phases can be found for $\Omega<\Omega_c$. For 
moderate values of the interaction, the ground 
state can occupy one of the two degenerate energy minima, thus acquiring a 
non-zero momentum and constant density. This dynamical regime is denominated 
plane-wave (PW) phase, and it is also characterized by the population imbalance of 
the ground state. More strikingly, when the interaction increases, the system 
enters the stripe phase, which shows density modulations as a result 
of the interference created by the occupation of the two momentum states  
\cite{Ho2011,Sinha2011}.

In what follows, we select a coherent coupling $\Omega$ in order for the system 
to be either in the single-minimum or the plane-wave regime, where the 
ground state presents a smooth density profile.
As we will see, in these regimes, the 3D Gross-Pitaevskii equation admits 
JV solutions as excited states. They present a localized depletion of 
the density along with an associated phase jump for each condensate component.
Their appearance can be understood by following a dynamical 
process in which, for a given $k_L$, the linear coupling $\Omega$ is 
adiabatically increased. Starting in the stripe phase regime, the length 
between density peaks increases with $\Omega$, up to some critical value 
$\Omega_c$, where dark-soliton-like structures evolve from the ground state of 
the system \cite{Cao2015}. Above the critical value, the solitonic structure 
can persist as an excited state in the other dynamical phases 
of the system, the SM or the PW phases. Previous works 
have demonstrated the existence and stability of such excited states for the 
different phases of a 1D system \cite{Achilleos2013}, resembling the features of 
corresponding states in scalar condensates. Our results demonstrate that, even 
in 1D systems, there exist additional solitonic configurations proper of 
two-component condensates. Besides, we report on the existence and 
stability of JVs in multidimensional settings with parameters typical of 
current experiments. In particular, we will focus on quasi-2D 
condensates produced by harmonic traps (with aspect ratio $\gamma=4$) sharing 
the transverse frequency $\omega_\rho=2\pi\times200$ Hz, a laser wavelength 
$\lambda_L=\sqrt{2}\pi/k_L=1064\,\mbox{nm}$, a coherent coupling 
$\Omega$ ranging from $10\, \mbox{kHz}$ (inside the PW phase) to $30\, 
\mbox{kHz}$ (inside the SM phase), and scattering lengths 
$a=101.41\,a_B$ and $a_{\uparrow\downarrow}=100.94\,a_B$, 
where $a_B$ is the Bohr radius, corresponding to the hyperfine states 
$|F=1,m_F=0\rangle$ and $|F=1,m_F=-1\rangle$ of $^{87}$Rb.

\section{Josephson vortices in 1D condensates}
\label{sec:1D}

The main properties of static JVs can be easily identified in 1D 
condensates, where analytic solutions to GPE are available in the absence of
SO coupling ($k_L=0$). Contrary to static dark solitons, JVs, being excited 
states with 
lower energy, have complex wave funtions that denote the presence of 
interspecies (spin) currents in the condensate.

\subsection{Without SO coupling}

It is instructive to start considering a two-component system with $k_L=0$. 
Then, the homogeneous 
($V_{trap}=0$) GPE (\ref{tdgpe}) admits two type of analytic stationary 
solitonic solutions (that have been addressed in the literature 
for the particular case $g_{\uparrow\downarrow}=0$ \cite{Kaurov2005,Qadir2012}), 
namely a dark soliton state
\begin{equation}
 \psi_{\uparrow,\downarrow}^{DS}(x)= \pm\sqrt{{n_T}}\, \tanh(x/\xi_\mu)\, ,
 \label{eq:DS}
\end{equation}
having a density-dependent healing length $\xi_\mu=\hbar/\sqrt{m\mu_{\rm 
eff}}$, with $\mu_{\rm eff}=\mu+\hbar\Omega/2$, and 
valid for all values of the coherent coupling $\Omega\ge0$, and 
a Josephson vortex state
\begin{equation}
 \Psi_{\uparrow,\downarrow}^{JV}(x) =  \sqrt{n_T}\left( 
\pm\tanh(x/\xi) + i 
\sqrt{1-\frac{2\hbar\Omega}{\mu_{\rm eff}}}\,\mbox{sech}(x/\xi)\right) \, ,
\label{eq:JV}
\end{equation}
with a limited range of existence $\mu_{\rm eff}>2\hbar\Omega$, and a healing 
length depending on the coherent coupling $\xi=\hbar/\sqrt{2m\hbar\Omega}$. In 
both states, the $\pm$ signs stand for the components $\uparrow$ and 
$\downarrow$ respectively, 
$n_T=n_\uparrow+n_\downarrow=\mu_{\rm eff}/(g+g_{\uparrow\downarrow})$ is the 
constant total density, there is no population imbalance 
$n_\uparrow=n_\downarrow=n_T/2$, and we have defined $\mu_{\rm eff}$ as an 
effective chemical potential  \cite{Gallemi2015b}, 
which is related to the condensate density. For every single value of $\mu_{\rm 
eff}$ there exists a unique DS state, but several JVs associated to different 
values of $\Omega$.

DS states in the absence of SO coupling, Eq. (\ref{eq:DS}), show the same 
symmetry between condensate components as the ground state, since from
$\Psi_\uparrow^{GS}=-\Psi_\downarrow^{GS}$ we get the corresponding 
spin-antisymmetric solitonic state $\Psi_\uparrow^{DS}=-\Psi_\downarrow^{DS}$. 
In the case of JVs, they fulfill 
$\Psi_\uparrow^{JV}=-(\Psi_\downarrow^{JV})^*$, which involves time 
reversal symmetry along with spin antisymmetry.
In the range of coexistence of DSs and JVs, the 
healing length linked to density is narrower than that associated to the 
coherent coupling, $\xi_\mu < \xi$, and, as a result, DSs possess higher 
energy ($\propto \hbar^2/m\xi_\mu^2$) than JVs. This fact affects their 
stability, making the JVs dynamically stable states all along their range of 
existence in 1D, and turning DSs into dynamically unstable ones whenever 
$\mu_{\rm eff}>2\hbar\Omega$. As we will see in the next sections, JVs can 
decay in multidimensional systems by snake instabilities, in the same manner 
as DSs do, in spite of the fact that the latter are still more energetic than 
the former.

\begin{figure}[tb]
\centering
\includegraphics[width=0.97\linewidth]{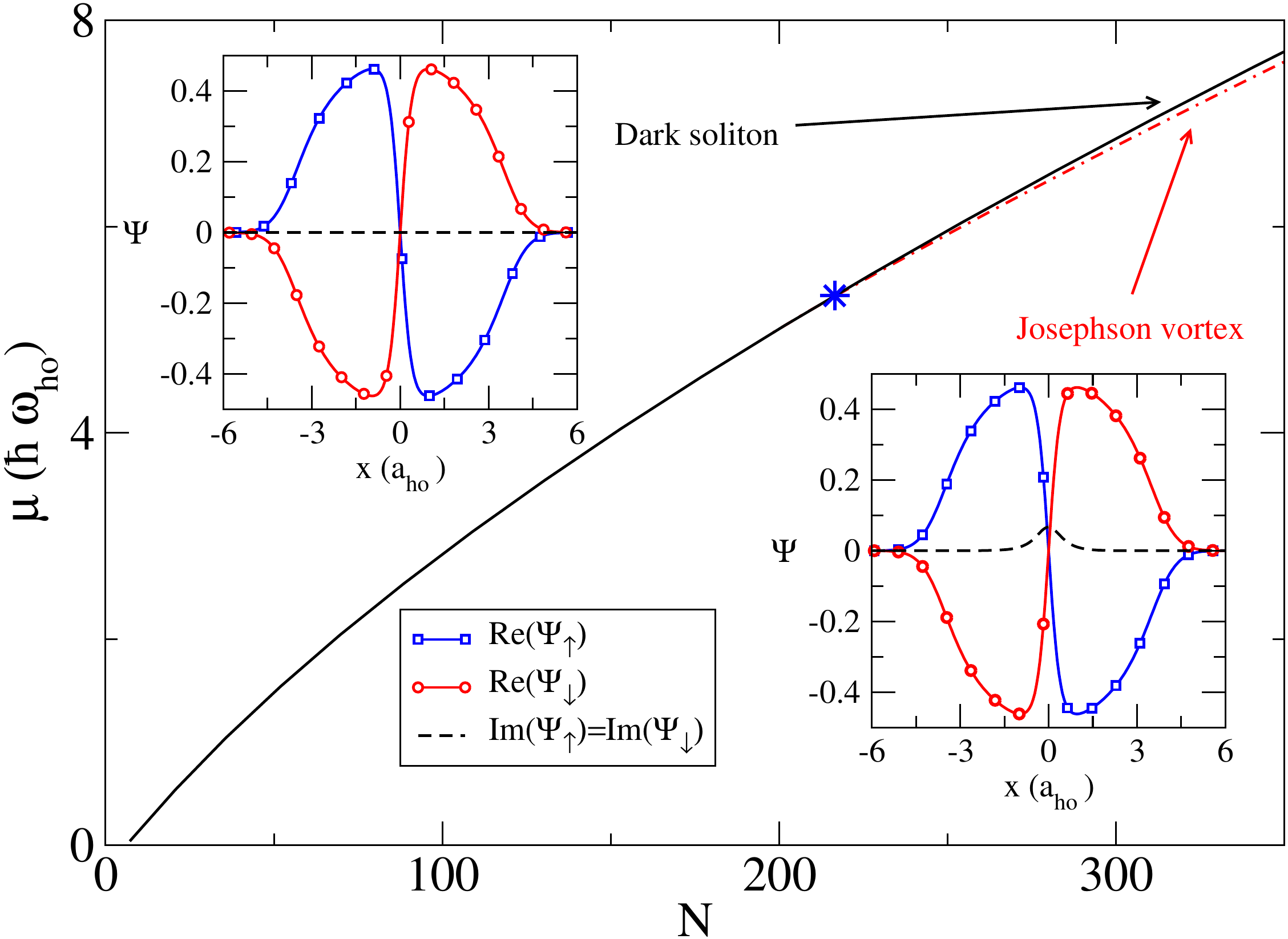}\\
\vspace{0.3cm}
\includegraphics[width=0.97\linewidth]{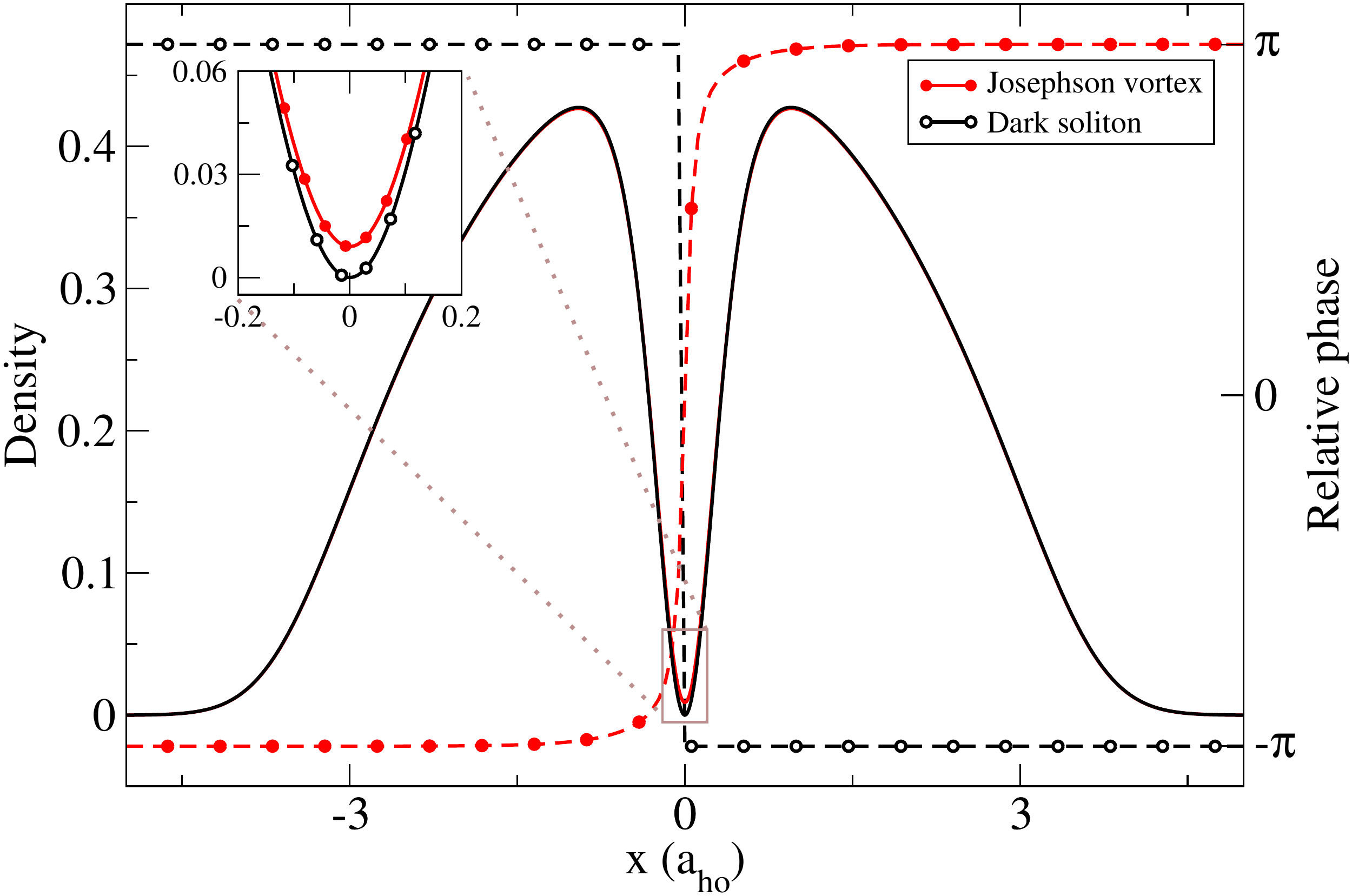}
\caption{Family of 1D DS and JV states for two-component 
condensates, in the absence of SO coupling, confined by a harmonic potential of 
frequency $\omega_{ho}=2\pi\times200$ Hz, and coupled by a Raman frequency  
$\Omega/\omega_{ho}=3.5$. Top panel: trajectories in the $\mu-N$ plane (main 
graph), and two nearby representative examples (insets, DS at the left
and JV at the right) corresponding to the (blue) star symbol. Bottom panel: 
total density (solid line) and relative phase (dashed line) of the DS (open 
circles) and the JV (filled circles). The inset zooms in the density depleted 
regions of both cases. }
\label{Fig:1D_DS_JV}
\end{figure}

The presence of a harmonic trapping potential does not modify qualitatively the 
type of solitonic solutions in two-component condensates without SO coupling. 
There also exist stationary states with the same symmetries as those discussed 
in the homogeneous case (Eqs. (\ref{eq:DS}) and (\ref{eq:JV})). The main 
difference arises from the continuation of such solitonic states from the 
noninteracting regime. Even in this linear case, the system admits both, 
(real) spin antisymmetric solutions $\Psi_\uparrow=-\Psi_\downarrow$:
\begin{equation}
 \psi^n_{\uparrow,\downarrow}(x)= \pm \, e^{-x^2/2a_{ho}^2}\, H_n(x/a_{ho})\, ,
 \label{eq:DStrap}
\end{equation}
where $H_n$ is the normalized Hermite polynomial of order $n$, and current 
states with the stronger symmetry $\Psi_\uparrow=-\Psi_\downarrow^*$:
\begin{equation}
 \psi^{n,m}_{\uparrow,\downarrow}(x)=  e^{-x^2/2\xi^2} 
\left(\, \pm\alpha H_n(x/\xi)+i \beta H_m(x/\xi)\, \right)\, ,
 \label{eq:JVtrap}
\end{equation}
where, $n\neq m$, and $\alpha$ and $\beta$ are real coefficients 
satisfying $\alpha^2+\beta^2=1$. While states corresponding to Eq. 
(\ref{eq:DStrap}) exist, with 
$\mu_{\rm eff}=\hbar\omega_{ho}(n+1/2)$, for every value of $\Omega$, the 
stronger 
symmetry of Eq. (\ref{eq:JVtrap}), having $\xi=a_{ho}$, requires that 
$\Omega=(n-m)\,\omega_{ho}$, so that $\mu_{\rm eff}=\hbar\Omega\,(2n+1)/2(n-m)$. 
We 
have checked that states (\ref{eq:DStrap}) and (\ref{eq:JVtrap}), which form a 
big manifold of excited states, persist, and even extend their range of 
existence, in the nonlinear regime. In particular, the nonlinear families of 
states starting from $\psi^1_{\uparrow,\downarrow}(x)= \pm \, 
e^{-x^2/2a_{ho}^2}\, H_1(x/a_{ho})$ and $\psi^{1,0}_{\uparrow,\downarrow}(x)= 
\pm \, e^{-x^2/2\xi^2}\left(\, \pm\alpha H_1(x/\xi)+i \beta H_0(x/\xi)\, 
\right)$ are the confined counterpart of the solitonic states Eq. 
(\ref{eq:DS}) and Eq. (\ref{eq:JV}), respectively. As an example, the top panel of Fig. 
\ref{Fig:1D_DS_JV} shows their trajectories in the $\mu$--$N$ plane for 
$\Omega/\omega_{ho}=3.5$, and two representative cases (in the insets) for an 
interaction $g_{1D}N/a_{ho}$= 30 $\hbar\omega_{ho}$, labeled by a blue star 
symbol, close to the bifurcation.

\begin{figure}[t!]
\centering
\includegraphics[width=0.97\linewidth]{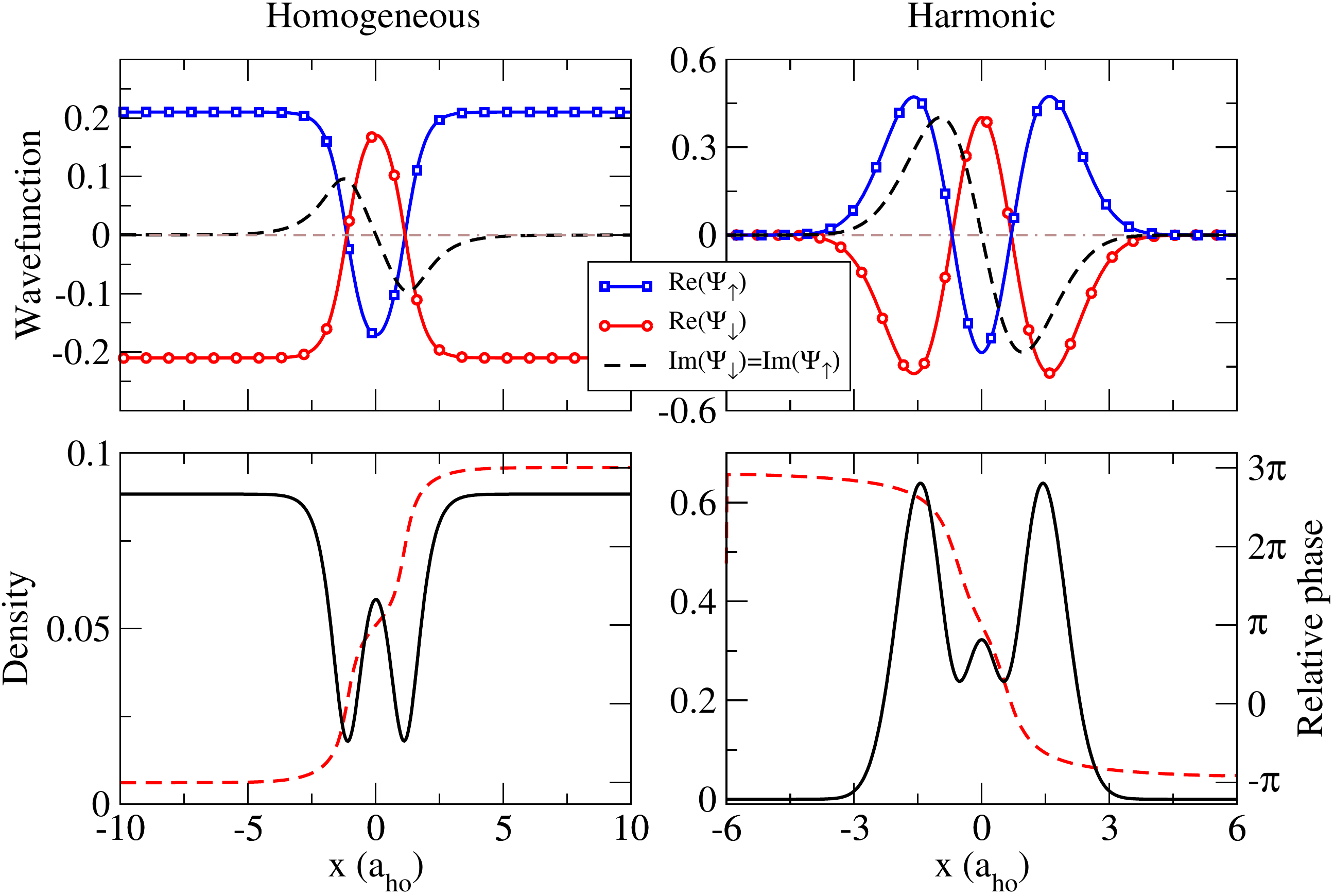}
\caption{Doubly charged JVs in homogeneous (left) and 
trapped (right) settings in the absence of SO coupling. Top panels show the 
condensate wave functions, and bottom panels depict the total densities (solid 
lines) and relative phases (dashed lines). For the sake of comparison, the same 
(harmonic oscillator) units have been used in all the graphs. Homogeneous case: 
$\mu=1.77\,\hbar\omega_{ho}$, $g_{\uparrow\downarrow}=0.2\,g$, and 
$\Omega/\omega_{ho}=0.7$. Trapped case: $\mu=2.22 \,\hbar\omega_{ho}$, 
$g_{\uparrow\downarrow}=0.9954\,g$, and $\Omega/\omega_{ho}=1.0$. }
\label{Fig:2JV}
\end{figure}

As can be seen in the graph of relative phase (bottom panel of 
Fig. \ref{Fig:1D_DS_JV}), JV states present a $2\pi$-jump
characteristic of the sine-Gordon solitons. In fact, it has been shown 
\cite{Son2002} that the linear equations in the relative phase, obtained by 
perturbing Eq. (\ref{tdgpe}), admits solutions containing domain walls 
(or \textit{kinks}) with a length scale that depends on the coherent coupling. 
The simplest of such solutions, made of a single kink, 
produces the referred $2\pi$-jump in the relative phase, whereas higher 
phase jumps can be introduced in the system by generating either several single
kinks, or a bound state of a 
kink and an antikink (\textit{breather}) \cite{Barone1982}. 
Recently, the dynamical generation of nonlinear excitations resembling the 
sine-Gordon breathers has been reported in coupled BECs \cite{Su2015}. 
Here, we show that doubly charged JVs are 
stationary solutions of the GPE (\ref{tdgpe}), which consist in a bound state 
of two kinks, making a doubly charged ($4\pi$-phase jump) soliton. 
They can be obtained by
introducing additional structure in the core of the JV, where by core we mean 
the spatial region of the condensate showing a depleted density (see Ref. 
\cite{Roditchev2015} for JV cores in superconductors). In terms of the 
singly charged JV (given by Eq. (\ref{eq:JV})), the core is enclosed in a 
healing length extension $\xi$ defined by the hyperbolic tangent shape of the 
wave function (in the homogeneous condensate), and is characterized by a 
density profile without zeros, described by the hyperbolic secant in the 
homogeneous case or the Hermite polynomial $H_0(x)$ in the (linear) trapped 
one. In contrast, a doubly charged JV state must present zeros in the imaginary 
part of the wave function inside the core, as 
it is the case of the examples represented in Fig. \ref{Fig:2JV}. They 
correspond to doubly charged JVs for homogeneous and trapped systems, obtained 
by numerically solving the 1D GPE (\ref{tdgpe}). Both states have no 
population imbalance. By direct observation of the wave function of the trapped 
condensate (top right panel), it follows that this 
state belongs to the family $\psi^{2,1}_{\uparrow,\downarrow}(x)= \pm \, 
e^{-x^2/2\xi^2}\left(\, \pm\alpha H_2(x/\xi)+i \beta H_1(x/\xi)\, \right)$.

\subsection{With SO coupling}

\begin{figure}[t!]
\centering
\includegraphics[width=0.97\linewidth]{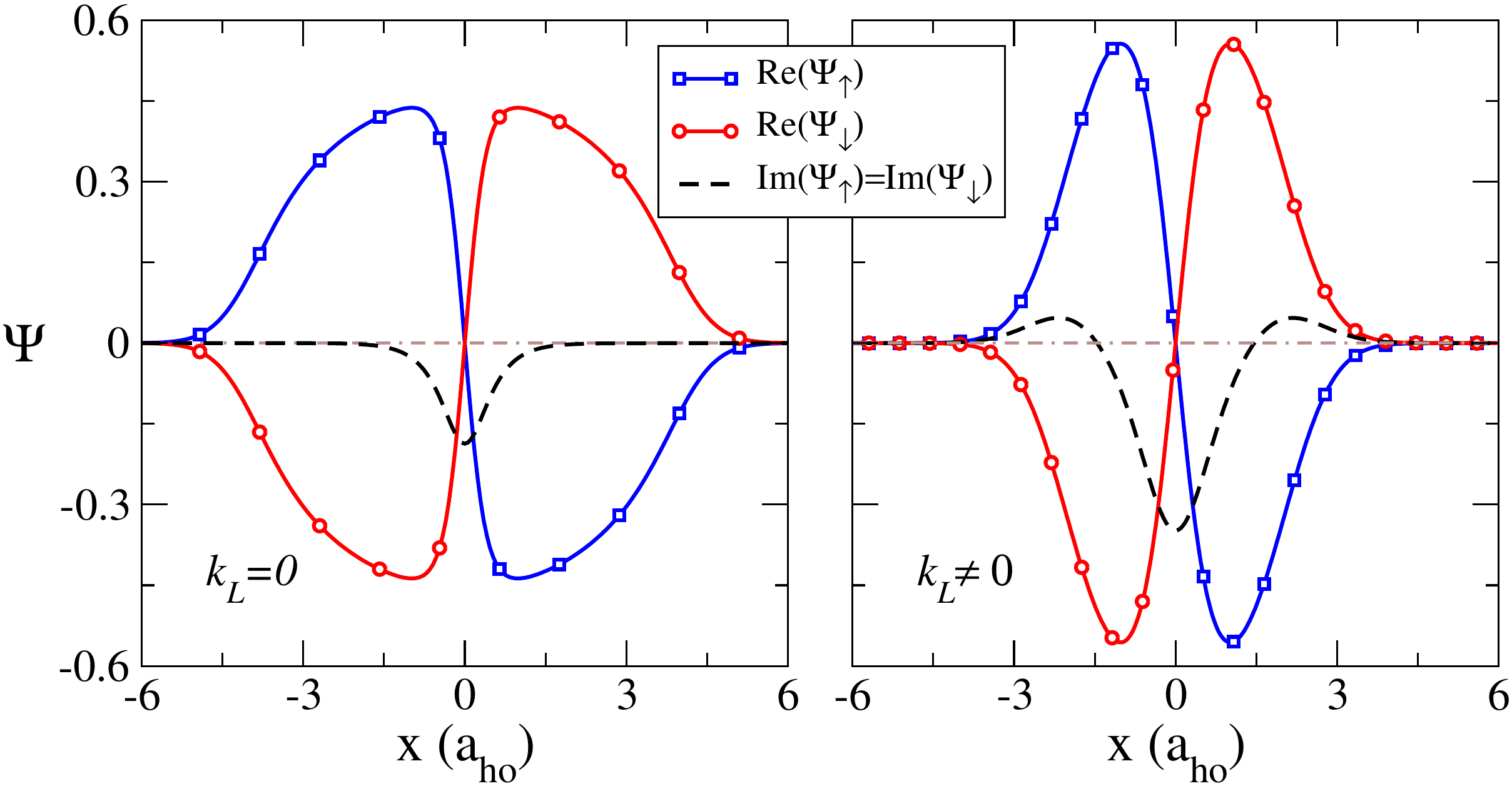}
\caption{JVs in the absence ($k_L=0$) and in the presence ($k_L\neq 0$) of SO coupling  
for trapped condensates with different parameters. Left: 
$k_L=0$, $\Omega/\omega_{ho}=3.5$, and $\mu=6.6 \,\hbar\omega_{ho}$. Right: 
$a_{ho}k_L=0.5$, $\Omega/\omega_{ho}=2$, and $\mu=1.7 \,\hbar\omega_{ho}$, within the SM phase. 
The interaction strengths are the same in both cases, 
$g_{\uparrow\downarrow}=0.9954\,g$. }
\label{Fig:1D_kl0_kl05}
\end{figure}

The laser coupling $k_L$ introduces a significant difference in 
the nature of the possible solitonic solutions to Eq. (\ref{tdgpe}). Dark 
solitons of the type described by Eq. (\ref{eq:DS}) are no longer possible, 
since their symmetry ($\Psi_\uparrow^{DS}=-\Psi_\downarrow^{DS}$) is not 
fulfilled by the equation of motion. Contrary to DSs, the symmetry of
JVs ($\Psi_\uparrow^{JV}=-(\Psi_\downarrow^{JV})^*$) is preserved by the SO 
coupling.
The wave functions of JVs, both in the homogeneous and trapped systems, are 
very similar to those shown in Fig. \ref{Fig:1D_DS_JV}. The main difference is 
the appearance of zeros (out of the soliton core) in their imaginary parts, as a 
result of the relative momentum introduced by $\pm \hbar k_L$. This feature can be 
seen in Fig. \ref{Fig:1D_kl0_kl05}, where we have plotted JVs in the absence of 
SO coupling, on the left, and in the presence of SO coupling, for 
$a_{ho}k_L=0.5$, on the 
right, for two different number of particles and coherent coupling. As it is 
reported below, multidimensional solitons present the same features (see Fig. 
\ref{Fig:3DJV}) discussed below.

\section{Josephson vortices in multidimensional, pancake-shaped condensates}
\label{sec:multiD}

Multidimensional BECs, composed of two coherently coupled (spin) components, 
can exhibit topological states in correspondence with the 1D ones presented 
before. In general, their existence depends on the interaction and on the 
inter-component coupling. As it is detailed below, our numerical results show 
that in harmonically trapped systems static DSs and JVs can be found in 
the absence of SO coupling ($k_L=0$), but only static JVs survives to non-zero 
values of
$k_L$. 

In order to find solitonic states, we evolve the full-3D Gross-Pitaevskii 
Eq.  (\ref{tdgpe}) in imaginary time, from the initial ansatz:
\begin{eqnarray}
\psi_\uparrow(\mathbf{r})=-\psi_\downarrow(\mathbf{r})=\chi_\perp(y,z)
\tanh(x/\xi(y,z)),
\label{ansatz}
\end{eqnarray}
where $\chi_\perp (x=0,y,z)$ is a transverse ground state, and 
$\xi(y,z)=\hbar/\sqrt{mg|\chi_\perp|^2}$ a local 
healing length. This ansatz, which corresponds to a dark soliton state in the 
$x$-direction (situated at $x=0$), follows from the approach proposed in Ref. 
\cite{MunozMateo2014} for scalar condensates, and, apart from being a real 
function, captures the main features of the solitonic solution (phase jump and 
density depletion) we are searching for. In particular, for the high-interaction 
regime it is explicitly defined through the Thomas-Fermi wave function 
$\chi_{TF}=\sqrt{(\mu_{\rm eff}-V(x=0,y,z))/(g+g_{\uparrow\downarrow})}$. In 
the case of 
solitons in the $y$-direction, $x$ and $y$ swap places in Eq. (\ref{ansatz}).
\begin{figure}[tb]
\centering
\includegraphics[width=0.97\linewidth]{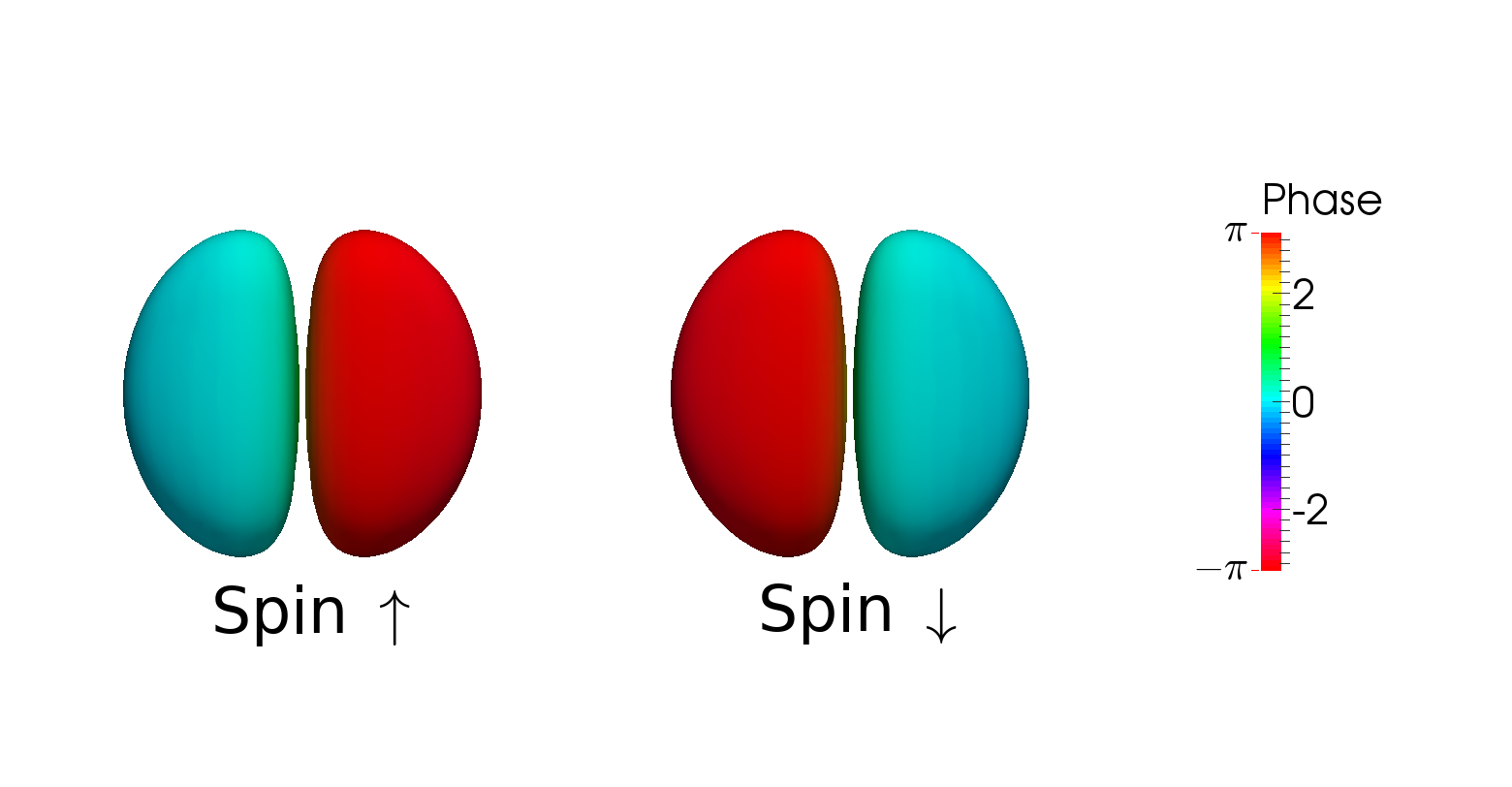}\\
\vspace*{-0.5cm}
\includegraphics[width=0.97\linewidth]{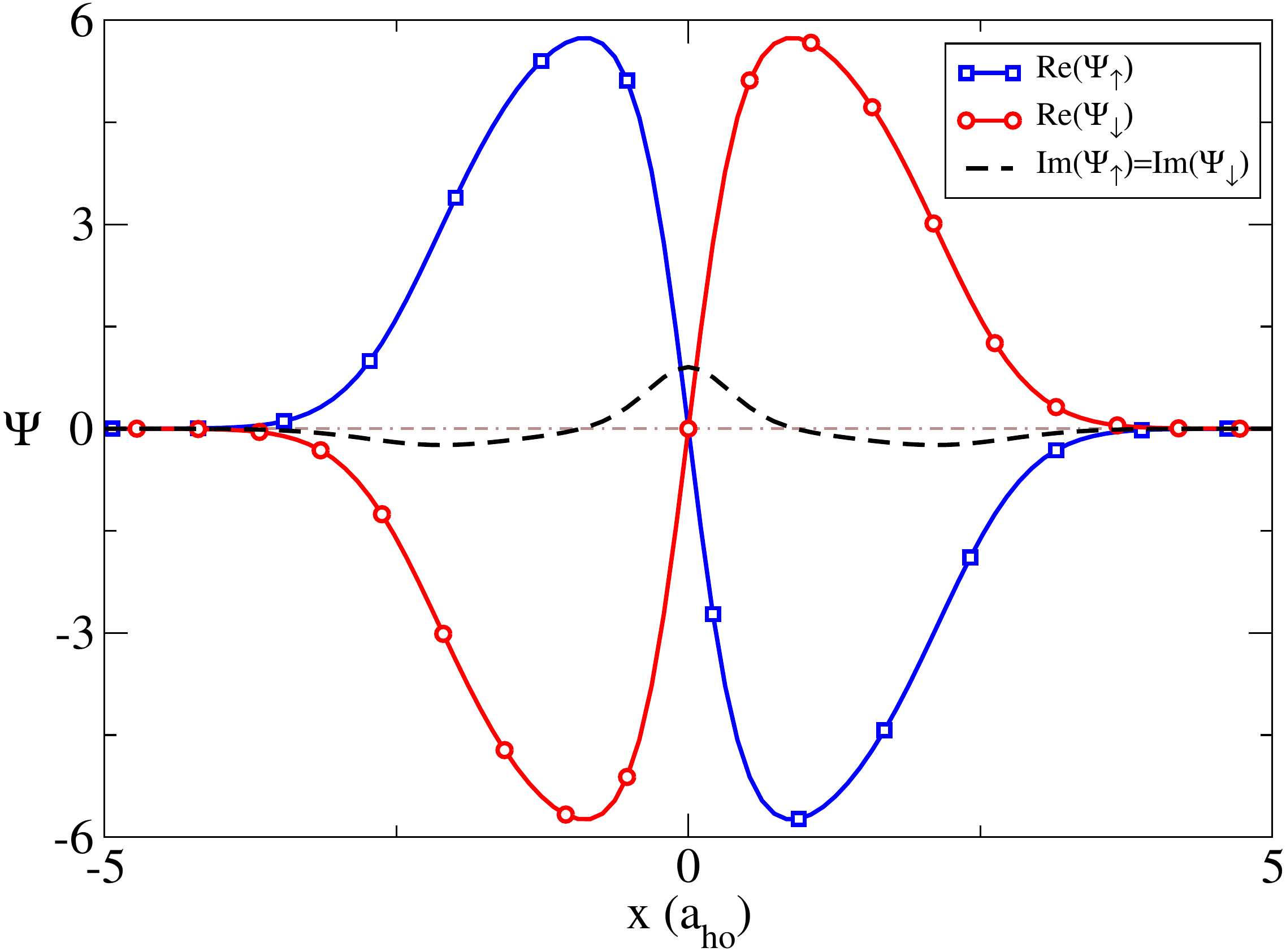}
\caption{3D Josephson vortex in a two-component condensate made of $N=500$ 
$^{87}$Rb atoms, trapped by a harmonic potential of frequency 
$\omega_\rho=2\pi\times 200$ Hz and aspect ratio $\gamma=4$,  with a spin-orbit 
coupling characterized by $\lambda_L=1064$ nm and $\Omega=30$ kHz. Top panel: 
density isocontours of the two components, at 5$\%$ of maximum density, 
coloured 
by phase. Bottom panel: real (solid) and imaginary (dashed) part of the 
wave functions (blue squares for spin-$\uparrow$ and red circles for 
spin-$\downarrow$) along the $x$-axis (for $y=z=0$). The imaginary part of the 
wave functions for both components are overlapped.}
\label{Fig:3DJV}
\end{figure}
Figure \ref{Fig:3DJV} shows a 3D JV in a spin-orbit-coupled system, obtained by 
following the procedure previously described. The condensate, which contains 
$N=500$ $^{87}$Rb atoms and is trapped by an external harmonic potential of 
frequency $\omega_\rho=2\pi\times 200$ Hz and aspect ratio $\gamma=4$, belongs 
to the SM phase, having the SO number $m\Omega/\hbar k_L^2=2.36$  
(with $\Omega=30$ kHz).
This is a stationary solution of the GPE (\ref{tdgpe}) with a non-trivial 
phase, given by a complex wave function (lower panel of Figure \ref{Fig:3DJV}) 
which involves the existence of inter-spin currents in the system. Comparing 
this state with the 1D JV without SO coupling, represented in Fig. 
\ref{Fig:1D_DS_JV}, 
the only difference comes from the zeros (out of the soliton core) in 
the imaginary part of the wave function.

\subsection{Stable Josephson vortices}

In the same way as it occurs with DSs in scalar BECs, it is possible to 
find dynamically stable JV states below an interaction threshold, or 
equivalently, below a critical value of the effective chemical 
potential $\mu_{\rm eff}$. Above this, JVs are unstable topological 
states, whose decay leads to the appearance of vortex lines in the system.

In order to demonstrate the stability of multidimensional JVs, we have 
analyzed SO coupled condensates in quasi-2D systems. To this aim, we have assumed
that the axial degrees of freedom of the 3D system are frozen, in 
such a way that the corresponding wave function is composed by an axial ($z$) 
separable part given by the ground state of the axial harmonic oscillator, 
that is, $\Psi(x,y,z)=\psi(x,y)\psi_z(z)$, where 
$\psi_z(z)=\exp(-z^2/2a_z^2)/\sqrt{a_z\sqrt{\pi}}$. After 
substituting this expression in Eq. (\ref{tdgpe}), multiplying 
by $\psi_z^*(z)$ and integrating over $z$, we 
get an equivalent two-dimensional GPE, where all the interaction strengths ($g$ 
and $g_{\uparrow\downarrow}$) are renormalized by the factor 
$a_{z}\,g_{2D}/g=1/\sqrt{2\pi}$.
\begin{figure}[tb]
\centering
\includegraphics[width=0.97\linewidth]{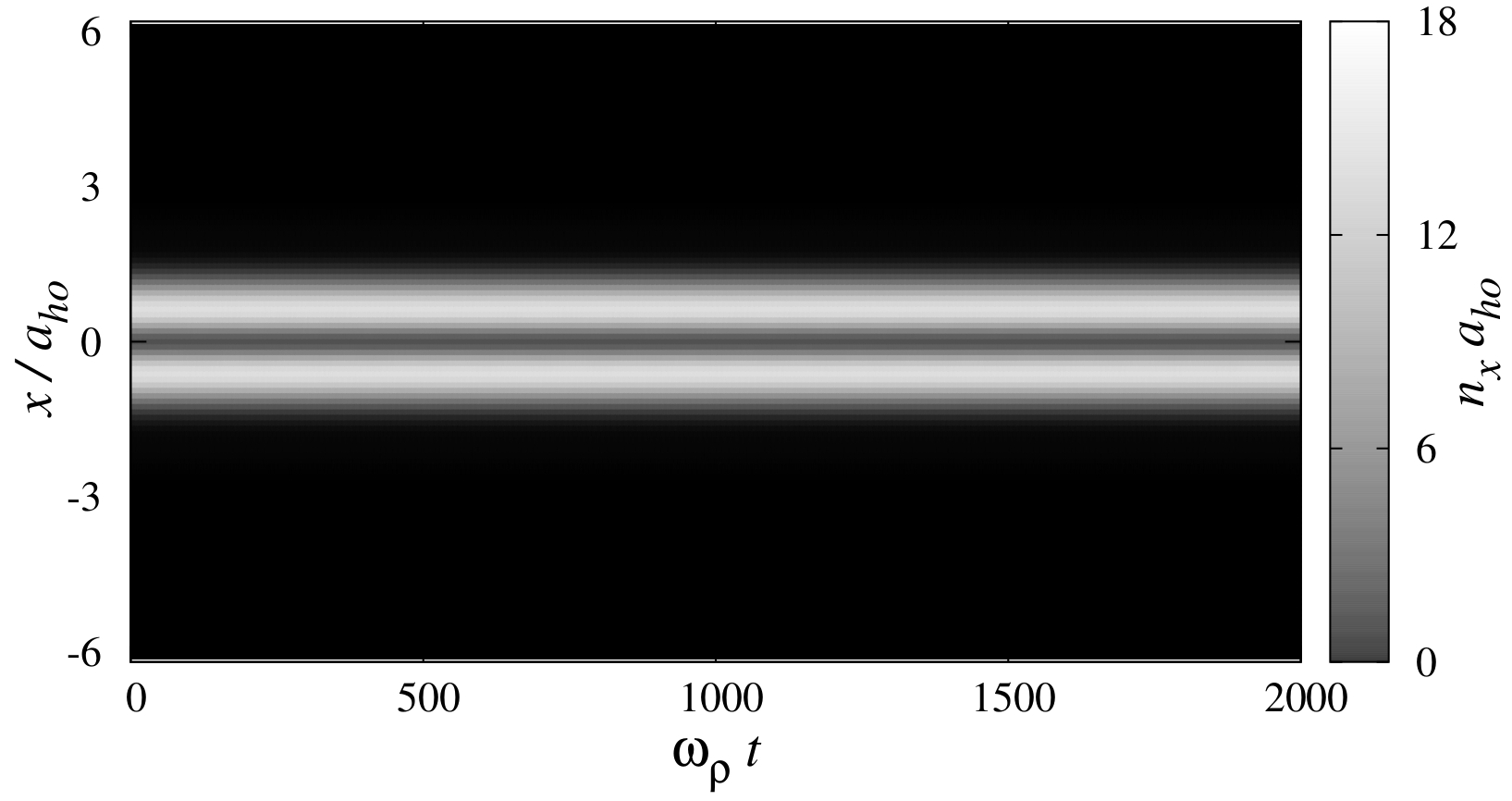}\\
\vspace*{-0.68cm}
\includegraphics[width=0.97\linewidth]{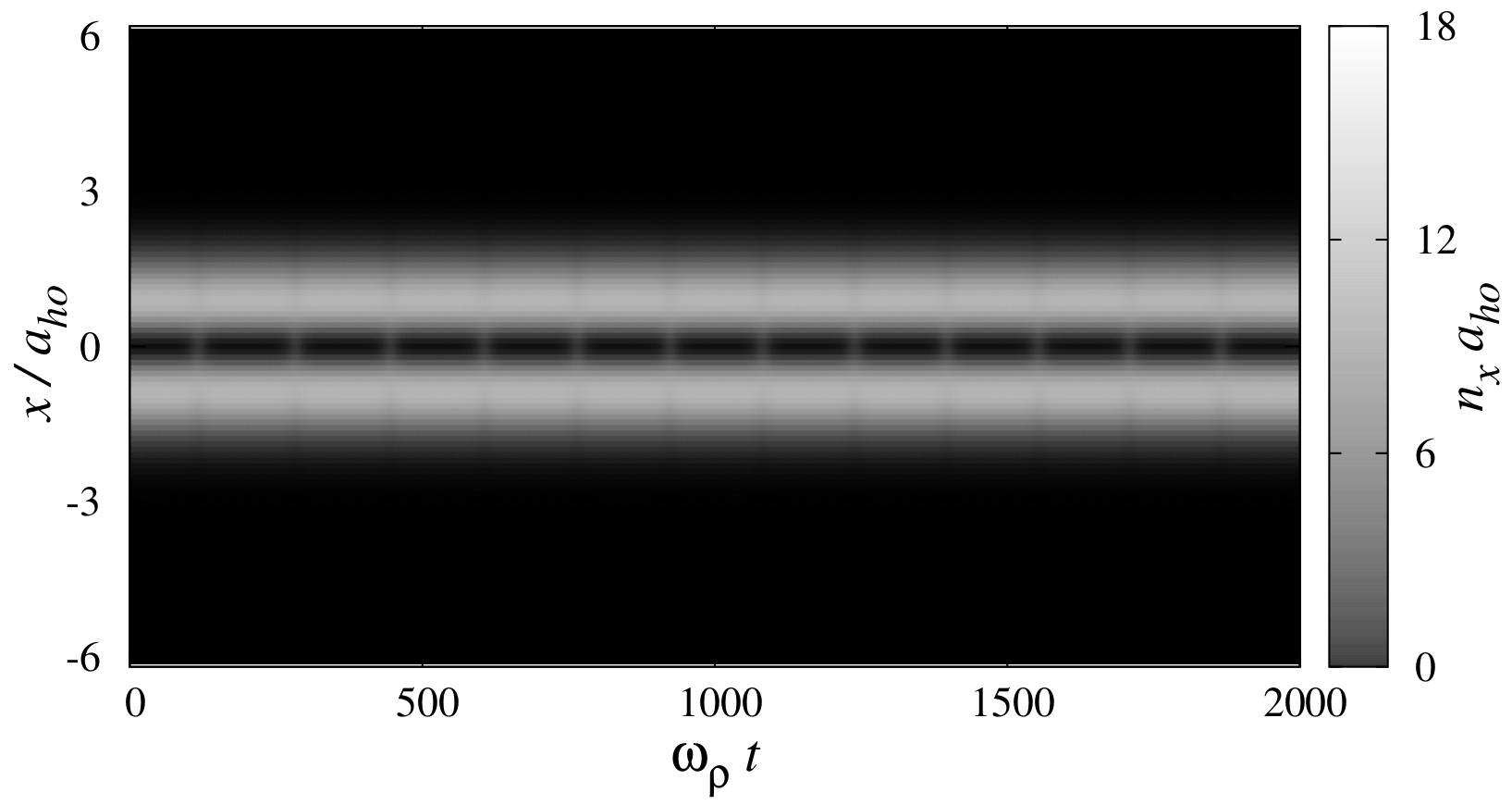}
\caption{The top (bottom) panel shows the real time evolution of a JV across 
the $x$-direction in
a 2D system with SO coupling, for $\Omega=10\,\mbox{kHz}$ 
($30\,\mbox{kHz}$) inside the the PW (SM) phase. Common parameters: $ 
N_{2D}=50$, 
$\omega_\rho= 2\pi\times200 \,\mbox{Hz}$, $\gamma=4$, and $\lambda_L=1064\, 
\mbox{nm}$.}
\label{Fig:evoJV_stable}
\end{figure}

Our results show that the critical value of $\mu_{\rm eff}$ for stability 
depends strongly on the presence of SO coupling, and specifically on the 
dimensionless number $m\Omega/\hbar k_L^2$. Without SO coupling, and for 
$\Omega=30 \,\mbox{kHz}$, JVs are stable in 2D condensates 
containing up to $N_{2D}=250$ particles. Under SO coupling, however, stability 
is 
dramatically reduced. Moreover, the richer phase diagram produces
differences on the stability for the same $\mu_{\rm eff}$, and provides 
different dynamical scenarios in the soliton decay.

In Fig. \ref{Fig:evoJV_stable} we show the real time evolution of JVs, across 
the $x$-direction, in the presence of SO coupling (with the 
wave vector $\mathbf{k}_L$ also lying on
$x$). For the initial state, at time $t=0$, we have added a Gaussian noise 
$\delta(\mathbf{r})$ to the stationary wave function 
$\psi(\mathbf{r})\rightarrow\psi(\mathbf{r})[1+0.01\delta(\mathbf{r})]$, that 
we will refer to as $1\%$-strength noise. The dimensionless density of the 
spin-$\uparrow$ component, integrated over the coordinate $y$ perpendicular to 
the soliton, $n_x(x)=\int |\psi_\uparrow(x,y)|^2 dy$, is represented against time, 
given in harmonic oscillator units.
The top panel of Fig. \ref{Fig:evoJV_stable}
corresponds to a stable JV in a 2D condensate containing $N_{2D}=50$ atoms, 
confined by a harmonic trap of frequency $\omega_\rho= 2\pi\times200\, 
\mbox{Hz}$ and aspect ratio $\gamma=4$. The SO number is $m\Omega/\hbar 
k_L^2=0.79$ (for $\Omega=10 \,\mbox{kHz}$), situating the system inside the 
PW phase. The horizontal dark line halving the atomic cloud marks the 
position of the JV, which remains unaltered during the whole evolution. 
The case represented in the bottom panel of Fig. \ref{Fig:evoJV_stable} is 
more peculiar. It has the same parameters as the condensate in the top panel 
except for a higher coherent coupling $\Omega=30 \,\mbox{kHz}$, which puts the system 
within the SM phase with a SO number $m\Omega/\hbar k_L^2=2.36$. 
Small periodic deviations from the initial (perturbed) state can be observed in 
the graph. Nevertheless, the stationary configuration is rapidly recovered, and 
the JV persists as a robust state.

\subsection{Decay dynamics}

\begin{figure}[t!]
\centering
\includegraphics[width=0.97\linewidth]{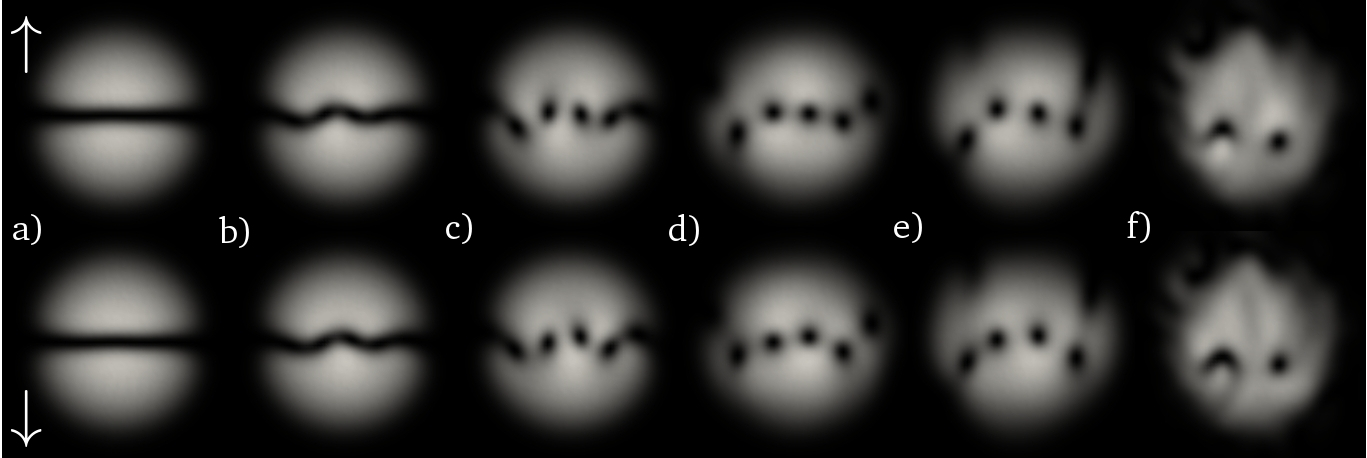}
\caption{Decay of a Josephson vortex, in the $y$-direction of a 
two-dimensional system with SO coupling, by snake instability of the soliton 
plane.
Condensate parameters: $ N_{2D}=2000$, $\omega_\rho= 2\pi\times200\,\mbox{Hz}$, 
$\gamma=4$, $\lambda_L=1064\, 
\mbox{nm}$, and $\Omega=30\,\mbox{kHz}$. The spin-$\uparrow$ (top) and 
spin-$\downarrow$ (bottom) density snapshots, 
from (a) to (f), correspond to times $\omega_\rho t = 3.5 , \; 
4.3, \; 5, \; 6, \; 8, \; \mbox{and} \; 20$, respectively. This 
sequence 
belongs to the real time evolution shown in Fig. \ref{Fig:evoJVy_SM}.}
\label{Fig:snake}
\end{figure}

Josephson vortices in multidimensional systems are unstable against transverse 
modes with long wavelengths. In this sense, the decay of 
JVs follows qualitatively that of DSs in scalar condensates, by snaking the 
soliton plane and producing vortex lines inside the system 
\cite{Kuznetsov1988}. In elongated condensates, the decay of 
a DS generates a solitonic vortex \cite{Brand2001,MunozMateo2014}, which 
survives as a dynamically stable state. However, in scalar disk-shaped condensates a 
single  vortex can not be generated after the soliton decay, because of angular 
momentum conservation; instead, a vortex dipole, constituted of a vortex and an 
antivortex, is left over \cite{Kuznetsov1995,Huang2003}. As we 
will demonstrate in what follows, this is also the case for disk-shaped 
SO coupled condensates. A vortex dipole per component is the remainder of the 
JV at the last stage of its decay. 

Figure \ref{Fig:snake} shows the whole time sequence of decay, from (a) to (f), for 
a JV across the $y$-direction in a $^{87}$Rb, 2D condensate containing 2000 
atoms. As can be seen, the two components follow a synchronized decay. After 
the snaking of the soliton plane, several pairs of vortex dipoles appear, 
almost overlapped, for each spin component. The vortices situated at the same 
position are connected by domain walls in the relative phase \cite{Son2002}. 
Finally, only one vortex dipole per component survives, and their subsequent  
evolution depends both on the particular dynamical regime of the system (i.e.  
either the SM or PW phase), and on the orientation of the JV plane with 
respect to the direction of the laser wave vector $\mathbf{k}_L$.

\begin{figure}[t]
\centering
\includegraphics[width=0.84\linewidth]{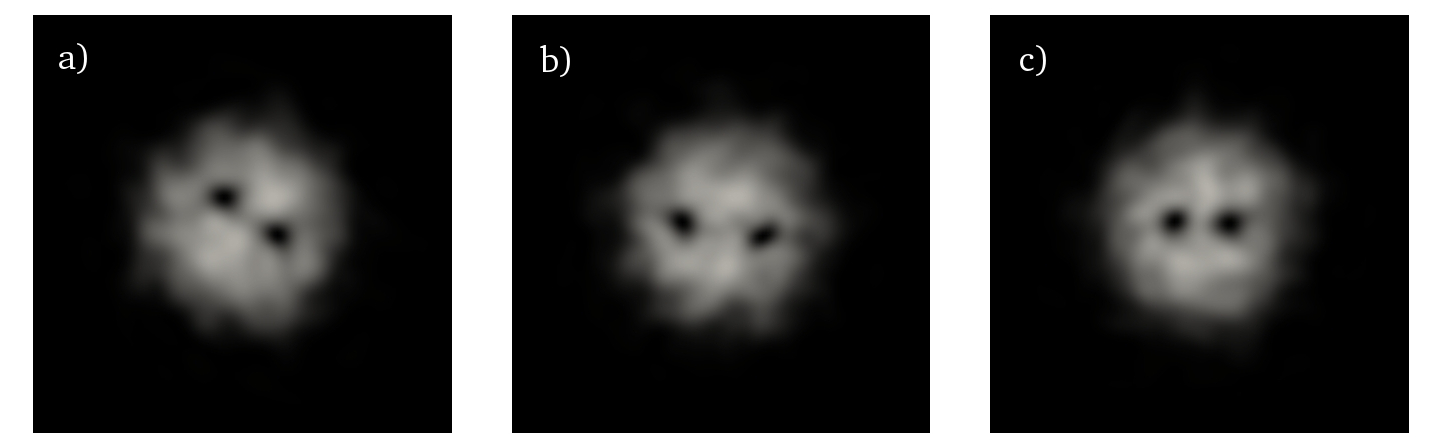}\\
\includegraphics[width=0.97\linewidth]{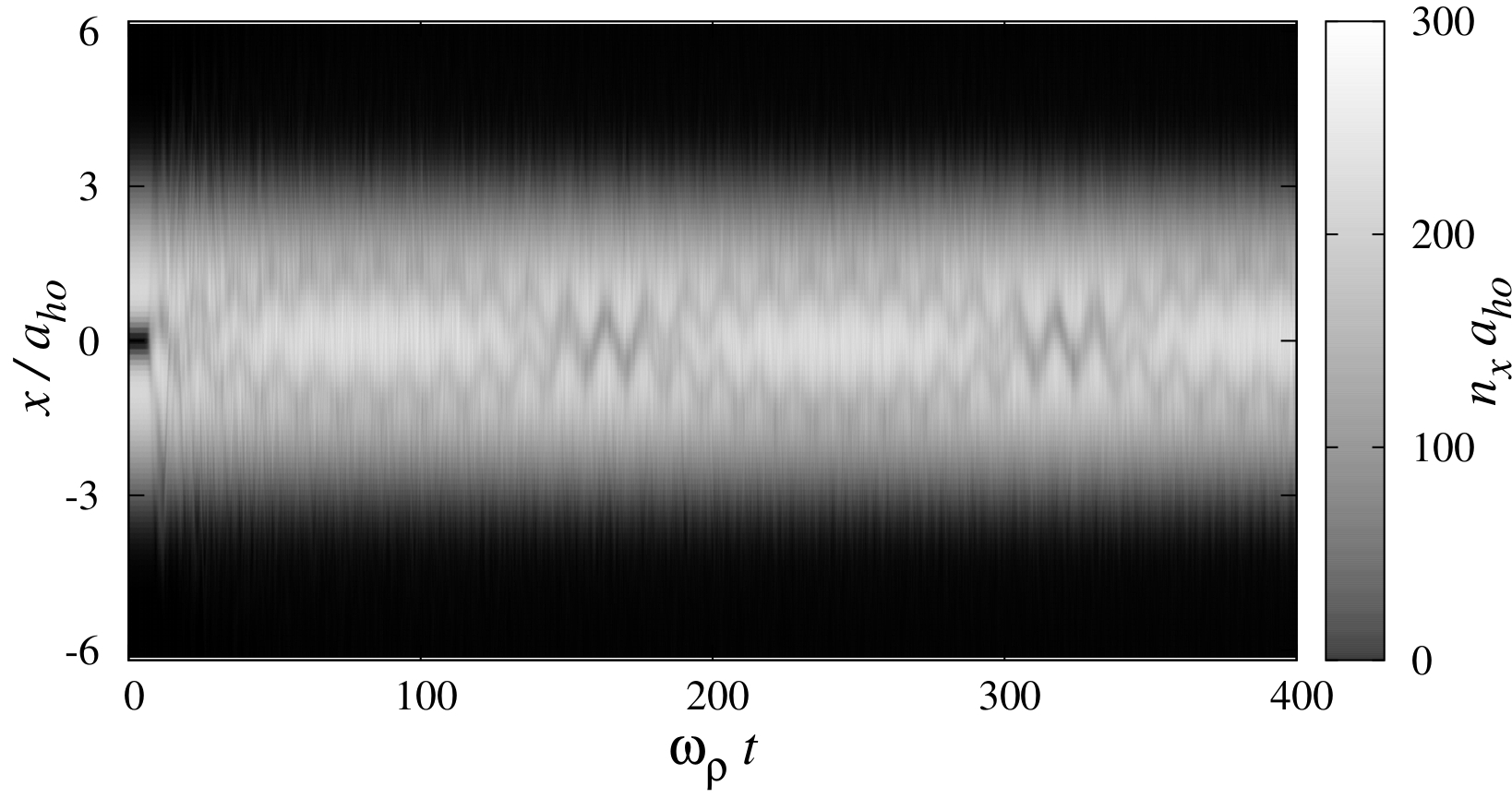}\\
\vspace*{-0.7cm}\hspace*{0.01cm}
\includegraphics[width=0.97\linewidth]{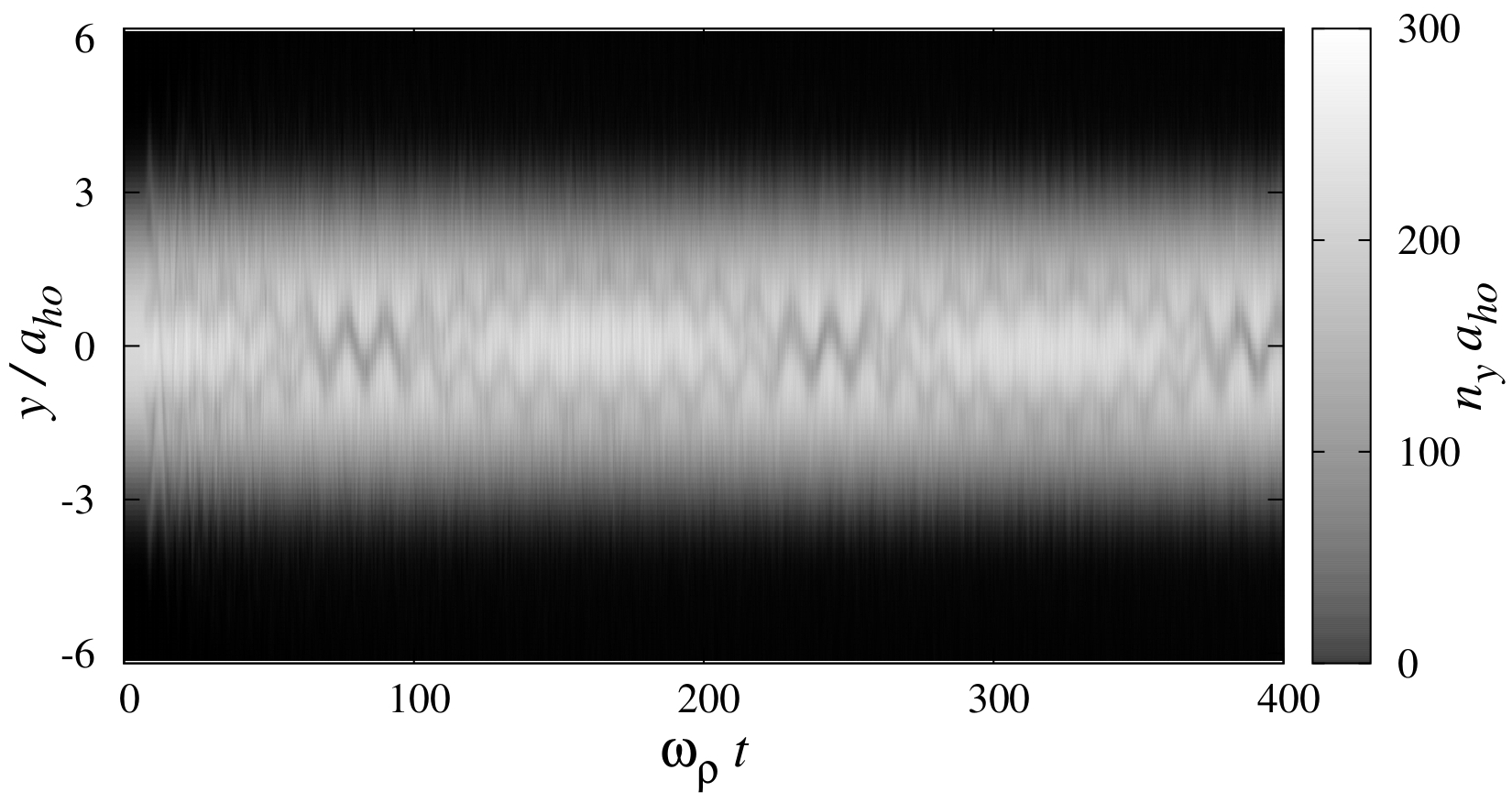}
\caption{Real time evolution of a Josephson vortex in the $x$-direction, in a 
2D condensate without SO coupling, and parameters: $ N_{2D}=2000$, 
$\omega_\rho= 2\pi\times200\,\mbox{Hz}$, $\gamma=4$, $\lambda_L=1064\, 
\mbox{nm}$, and 
$\Omega=10\,\mbox{kHz}$. Upper panel: spin-$\uparrow$-density snapshots, from 
(a) to (c), at times $\omega_\rho t = 51 , \; 73, \; \mbox{and} \; 78$, 
respectively. Middle and lower panels: non-dimensional spin-$\uparrow$ density integrated 
over the $y$ (middle) and $x$ (bottom) directions.}
\label{Fig:evoJV}
\end{figure}

For the sake of comparison, we consider first the decay dynamics in the 
absence of SO coupling. In Fig. \ref{Fig:evoJV} we show the real time evolution 
of a 
stationary JV after adding a 1$\%$-strength noise. The two lower panels 
represent the dimensionless density $a_{ho} n_x(x)$ and $a_{ho} n_y(y)$ for one 
condensate component (spin-$\uparrow$).
After few $\omega_\rho t$ cycles, the JV (thick black line) decays into a 
vortex dipole (upper snapshots of Fig. \ref{Fig:evoJV}) that describes a zigzag 
path (one per vortex) in the $x$ and $y$ coordinates. The dynamics of the 
vortex dipole consists in a superposition of a rotation around the center of 
the trap (low frequency oscillation in the graph), due to the rotational 
symmetry of the condensate, and a vibration of the relative position of the 
vortices (high frequency oscillation).

\begin{figure}[t!]
\centering
\includegraphics[width=0.97\linewidth]{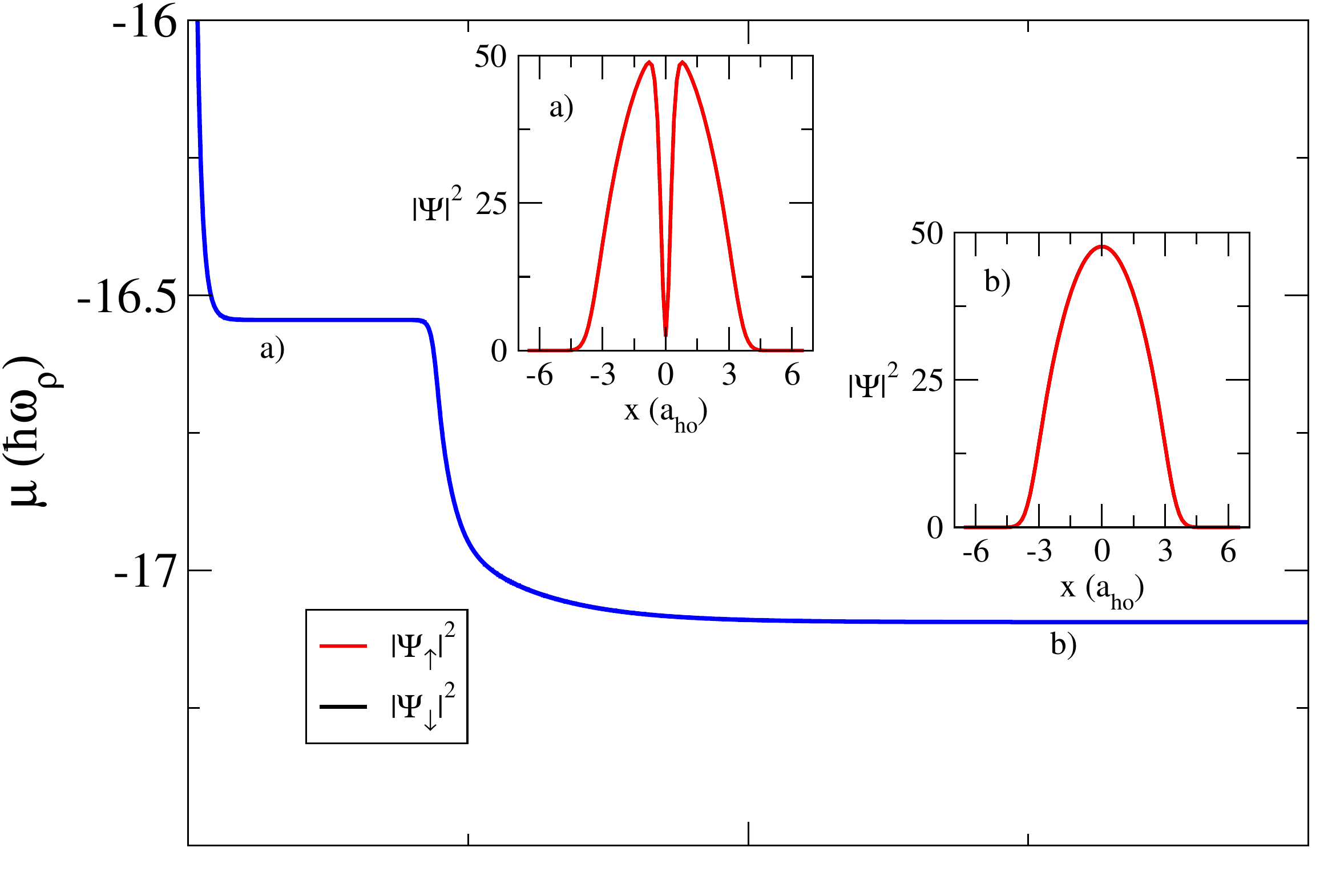}\\
\vspace*{-3mm}
\includegraphics[width=0.97\linewidth]{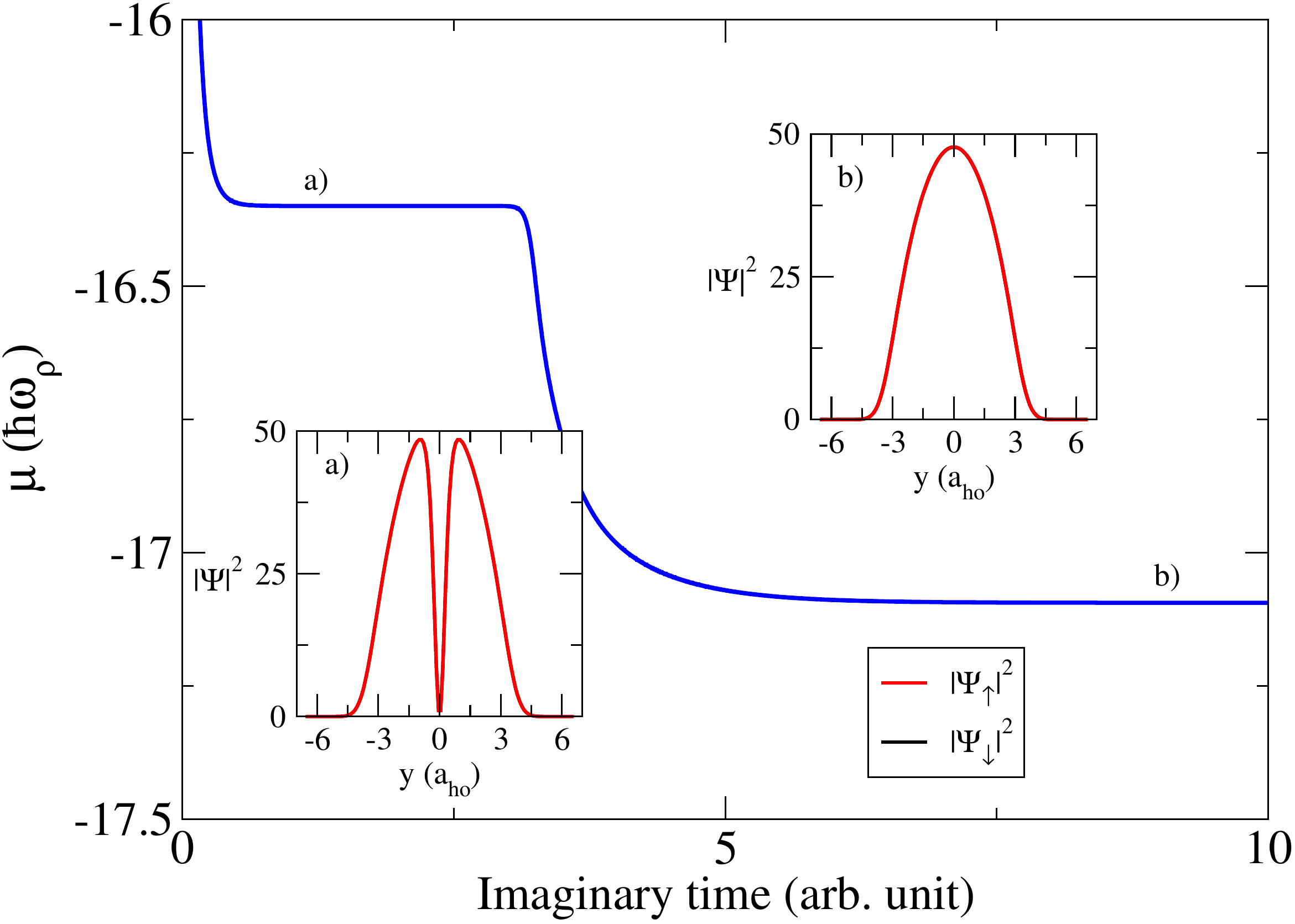}
\caption{Imaginary time evolution of JVs, in the $x$ (top) and $y$ (bottom) 
directions, within the SM phase, with parameters: $N_{2D}=2000$, 
$\omega_\rho=2\pi\times200\,\mbox{Hz}$, $\gamma=4$, 
$\lambda_L=1064\,\mbox{nm}$, 
$\Omega=30 \,\mbox{kHz}$. The insets, showing the spin densities along 
selected 
axis, correspond to different stationary states, from the initial JV up to the 
final ground state. }
\label{Fig:imagJV_SM}
\end{figure}

When SO coupling is switched on, the rotational symmetry is broken, and the 
vortex 
dipole that appears after the decay of the JV can not rotate freely. 
The term $-\mathbf{\hat p}\Psi \cdot \hbar \mathbf{k}_L$ enters 
the GPE (\ref{tdgpe}) as the potential of a magnetic dipole 
$-{\boldsymbol \mu}_S \cdot \mathbf{B}$, where ${\boldsymbol \mu}_S$ is the 
magnetic moment of a spin-$1/2$ particle and $\mathbf{B}$ is the static 
magnetic field. 
In this way, the velocity of the vortex dipole $-\mathbf{\hat 
p}\psi_\uparrow$, which is perpendicular to the dipole direction, interacts 
with the SO momentum $\mathbf{k}_L$, generating an aligning torque given by 
$\mathbf{\hat p}\psi_\uparrow \times 
\hbar \mathbf{k}_L$. Therefore, the orientation of the JV is crucial for the 
later dynamics of the emergent vortex dipoles, since their alignment follow 
the JV one. Given that this orientation and the velocity field of the dipoles 
are the same for both components, the opposite laser momentum $\pm \hbar k_L$, yields 
to energies and acting torques with different signs. For this reason, 
in what follows, we address the decay 
of JVs in different dynamical regimes, and with different orientations of the 
soliton plane, along the laser wave vector ($x$) and perpendicular to it ($y$).

\subsubsection{Josephson vortices in the SM phase}

\begin{figure}[t!]
\centering
\includegraphics[width=0.84\linewidth]{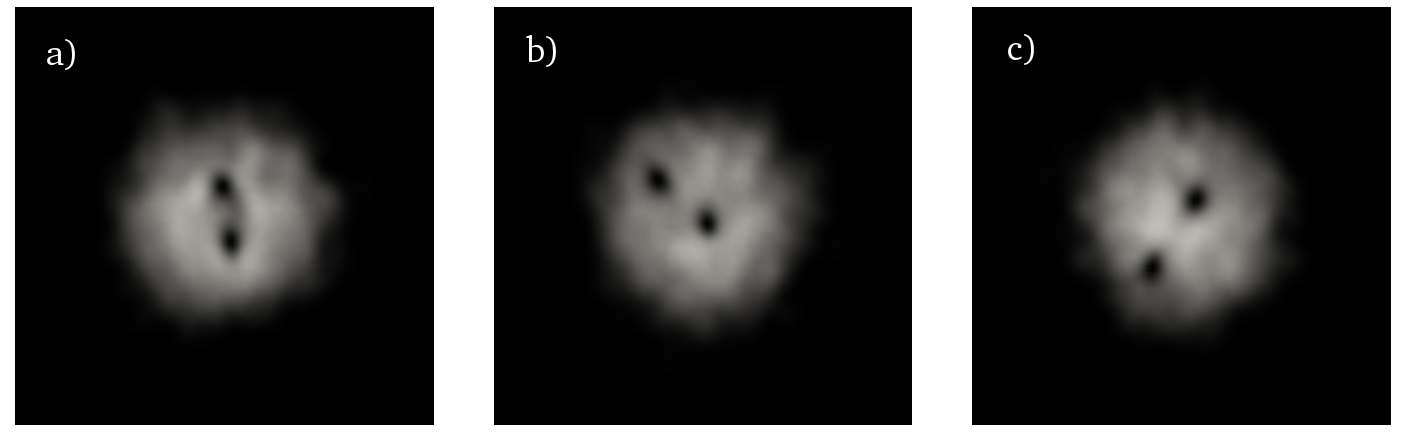}\\
\includegraphics[width=0.97\linewidth]{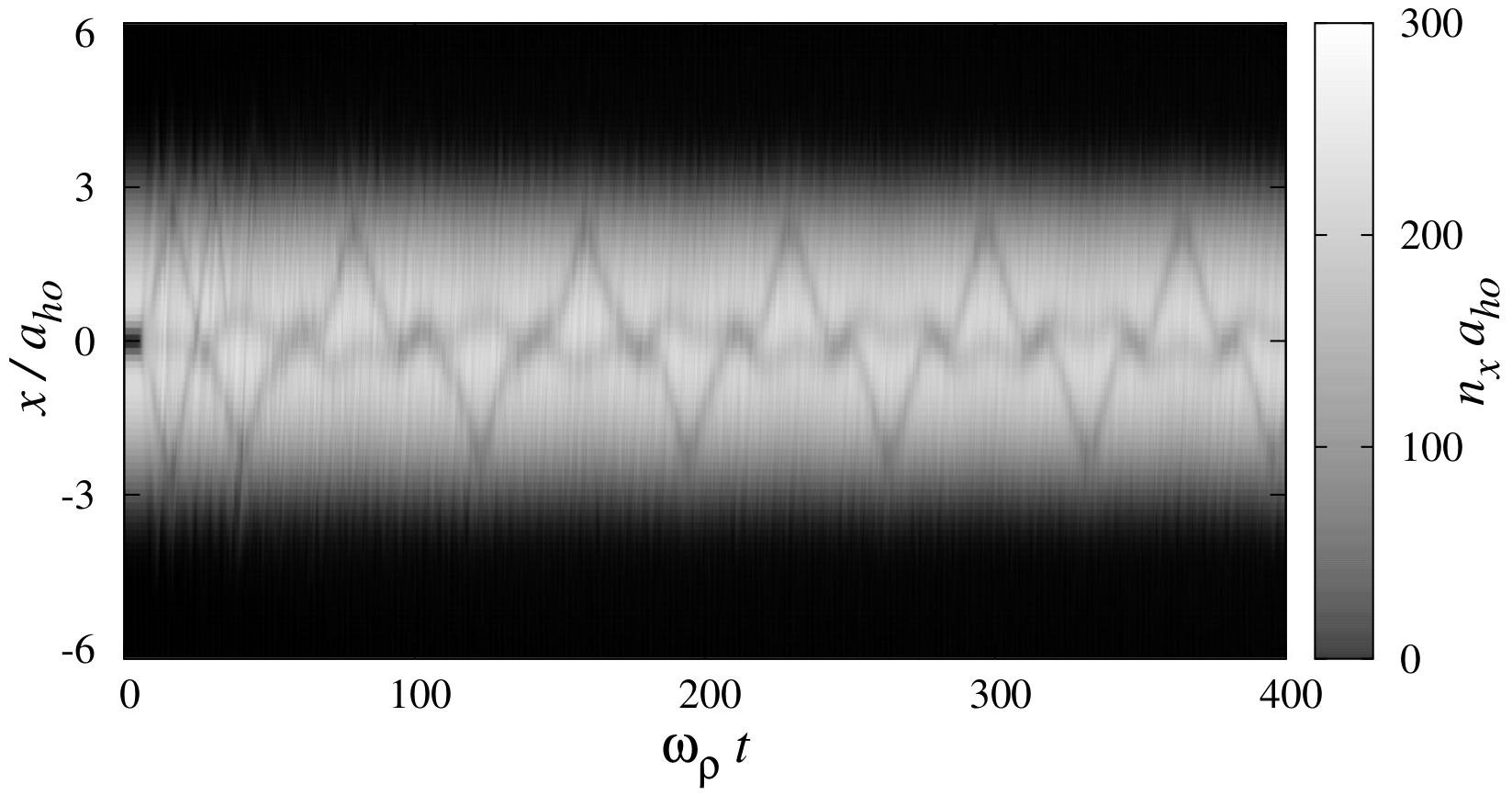}\\
\vspace*{-0.69cm}
\includegraphics[width=0.97\linewidth]{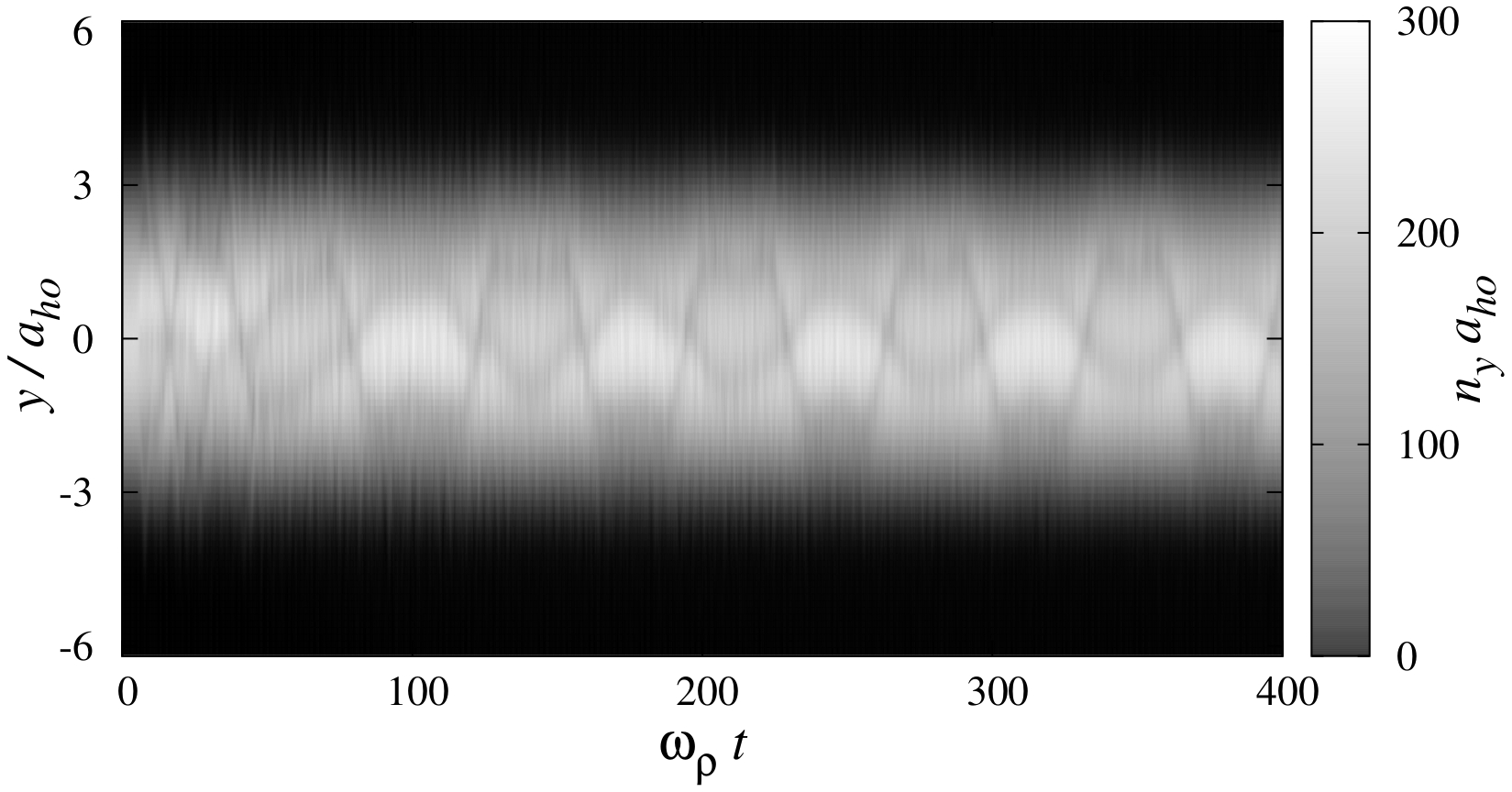}
\caption{ Real time evolution of the Josephson vortex, in the $x$-direction, 
shown in the top panel of Fig. \ref{Fig:imagJV_SM}. Upper panel: 
spin-$\uparrow$-density snapshots, from (a) to (c), at times $\omega_\rho t = 
99 , \; 112, \; \mbox{and} \; 124$, respectively. Lower panels: 
evolution of the non-dimensional spin-$\uparrow$ density integrated 
over the $y$ (top) and $x$ (bottom) directions.}
\label{Fig:evoJVx_SM}
\end{figure}

In the SM phase, the ground state of the system does not present population 
imbalance and the density profiles of both condensate components coincide. 
This is also the case for the excited states containing JVs. As can be seen 
in  Fig. \ref{Fig:imagJV_SM}, the imaginary-time evolution of our initial 
ansatz Eq. (\ref{ansatz}) reaches the energy plateaus of JVs, represented 
in the insets Fig. \ref{Fig:imagJV_SM}(a), for the two orientations 
considered, $x$ in the upper panel of the figure and $y$ in the lower one, 
before falling to the ground state of the system, depicted 
in the insets Fig. \ref{Fig:imagJV_SM}(b). During the whole evolution, obtained 
for the coherent coupling $\Omega=30\,\mbox{kHz}$ and SO number $m\Omega/\hbar 
k_L^2=2.36$, the population imbalance remains at zero.

JV states have a nonzero mean momentum. Therefore, their mean energy  
depends on their orientation because of the term associated to the SO coupling. 
As a 
result, for the same parameters, a JV across the $y$-direction has higher energy 
than a JV across $x$ (as can be seen in the main graphs of Fig. 
\ref{Fig:imagJV_SM}). In the latter case, the momentum originated by the JV is 
aligned with the laser wave vector, and then the energy of the system is 
reduced. On the other hand, for the JV across the $y$-direction the SO coupling 
has no 
influence on the mean energy.

\begin{figure}[t!]
\centering
\includegraphics[width=0.84\linewidth]{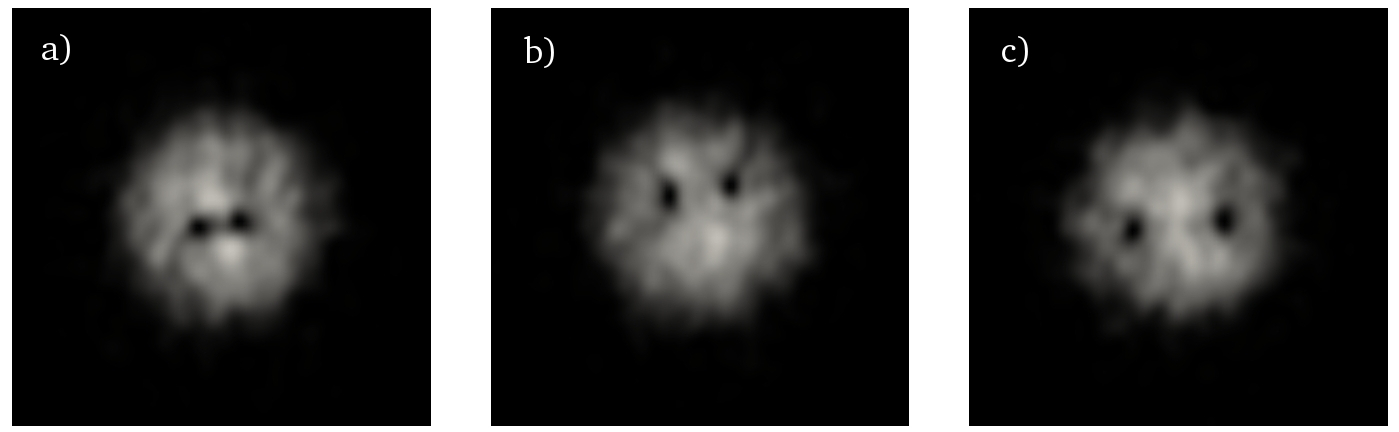}\\
\includegraphics[width=0.97\linewidth]{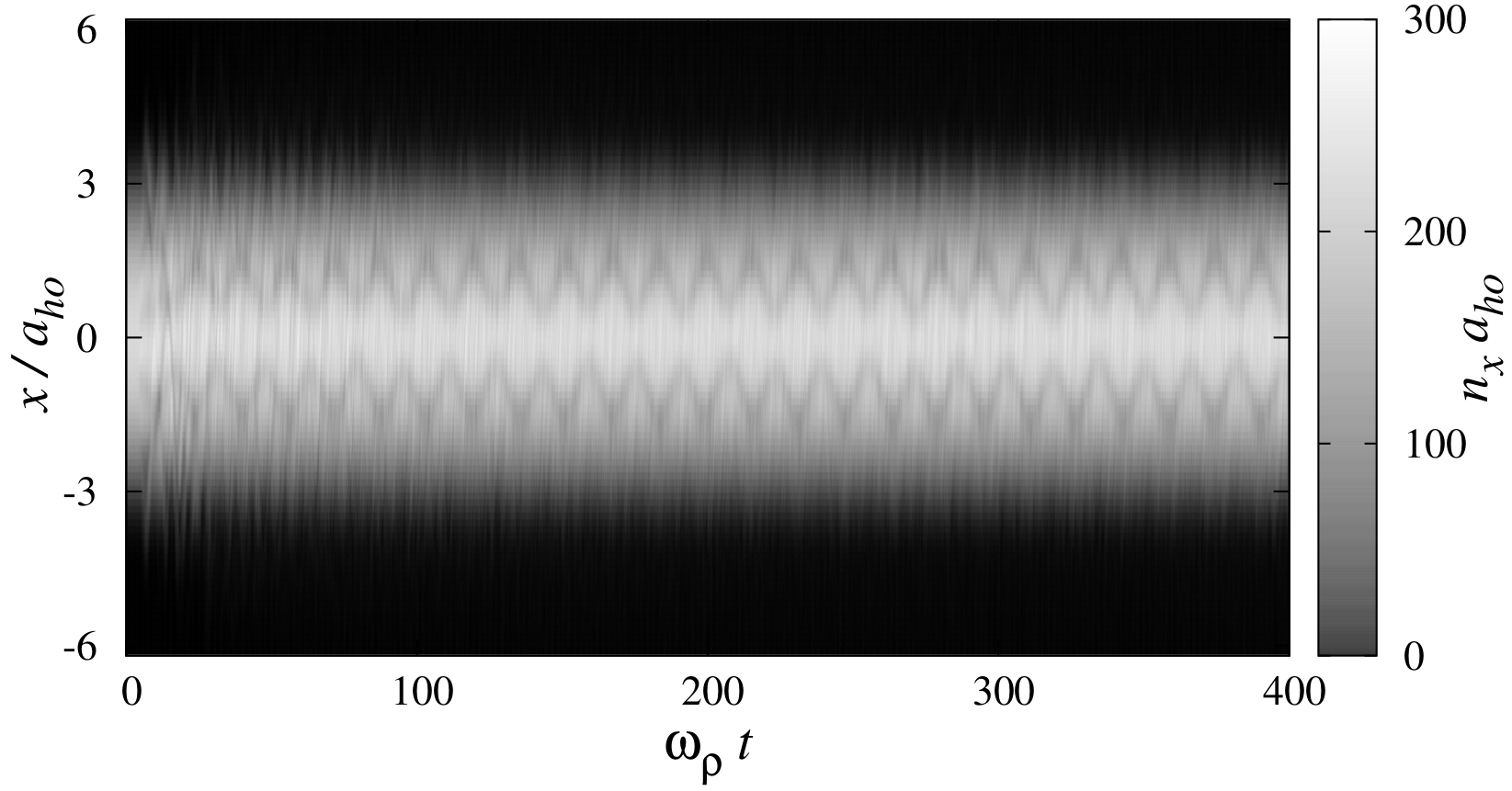}\\
\vspace*{-0.68cm}
\includegraphics[width=0.97\linewidth]{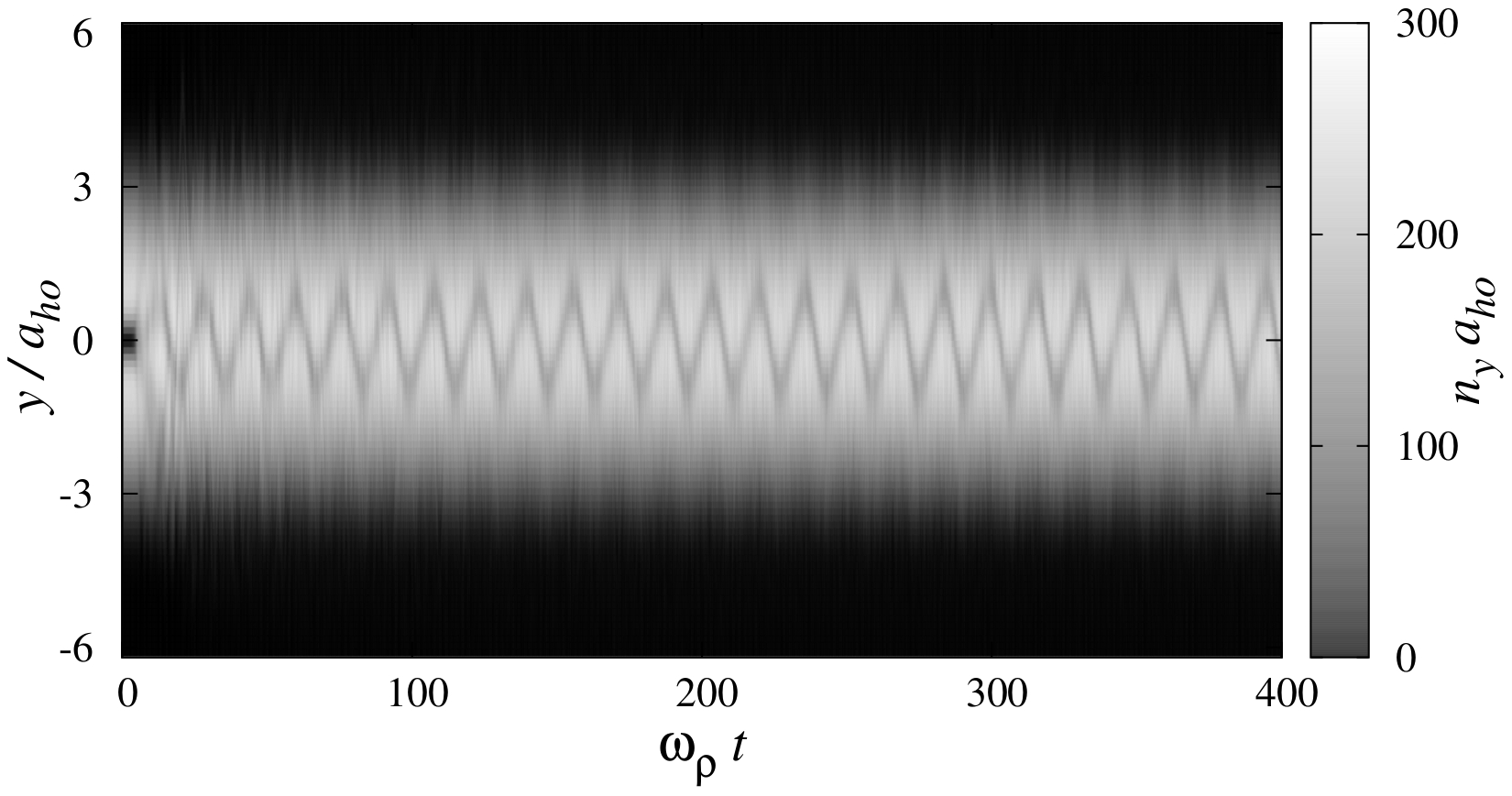}
\caption{ Same as Fig. \ref{Fig:evoJVx_SM}, for the JV, in the $y$-direction, 
shown in the bottom panel of Fig. \ref{Fig:imagJV_SM}. The snapshots, from (a) 
to (c), correspond to times $\omega_\rho t = 102 , \; 108, \; \mbox{and} \; 
112$, respectively.}
\label{Fig:evoJVy_SM}
\end{figure}

In order to analyze the dynamics of the JVs found in Fig. \ref{Fig:imagJV_SM}, 
we have performed their real time evolution by numerically solving the GPE 
(\ref{tdgpe}), after adding a 1$\%$-strength noise to the respective stationary 
states. The results are plotted in Fig. \ref{Fig:evoJVx_SM} for the JV oriented 
across $x$, and Fig. \ref{Fig:evoJVy_SM} for the JV oriented across $y$.
In the former case the velocities of the emerging vortex dipoles are aligned 
with the laser wave vector. One of the dipoles lays on the 
energy minimum of the SO potential, whereas the other one occupies the energy 
maximum. This configuration is unstable and degenerate against exchange of the 
positions of the vortex and the anti-vortex. As a consequence, the system 
oscillates between one configuration and the other. The whole picture can be 
seen in the evolution of Fig. \ref{Fig:evoJVx_SM}, where the peaked paths 
correspond to the motion of a vortex around the anti-vortex in order to 
exchange their positions.
When the JV is oriented across $y$ (Fig. \ref{Fig:evoJVy_SM}) the evolution 
resembles the case without SO coupling, except by the lacking of rotational 
motion of 
the dipoles around the center of the trap, because of the absence of rotational 
symmetry. The vortex dipoles are vibrating along with a small precession, 
keeping the orientation of the dipoles.

\subsubsection{Josephson vortices in the PW phase}

\begin{figure}[t!]
\centering
\includegraphics[width=0.97\linewidth]{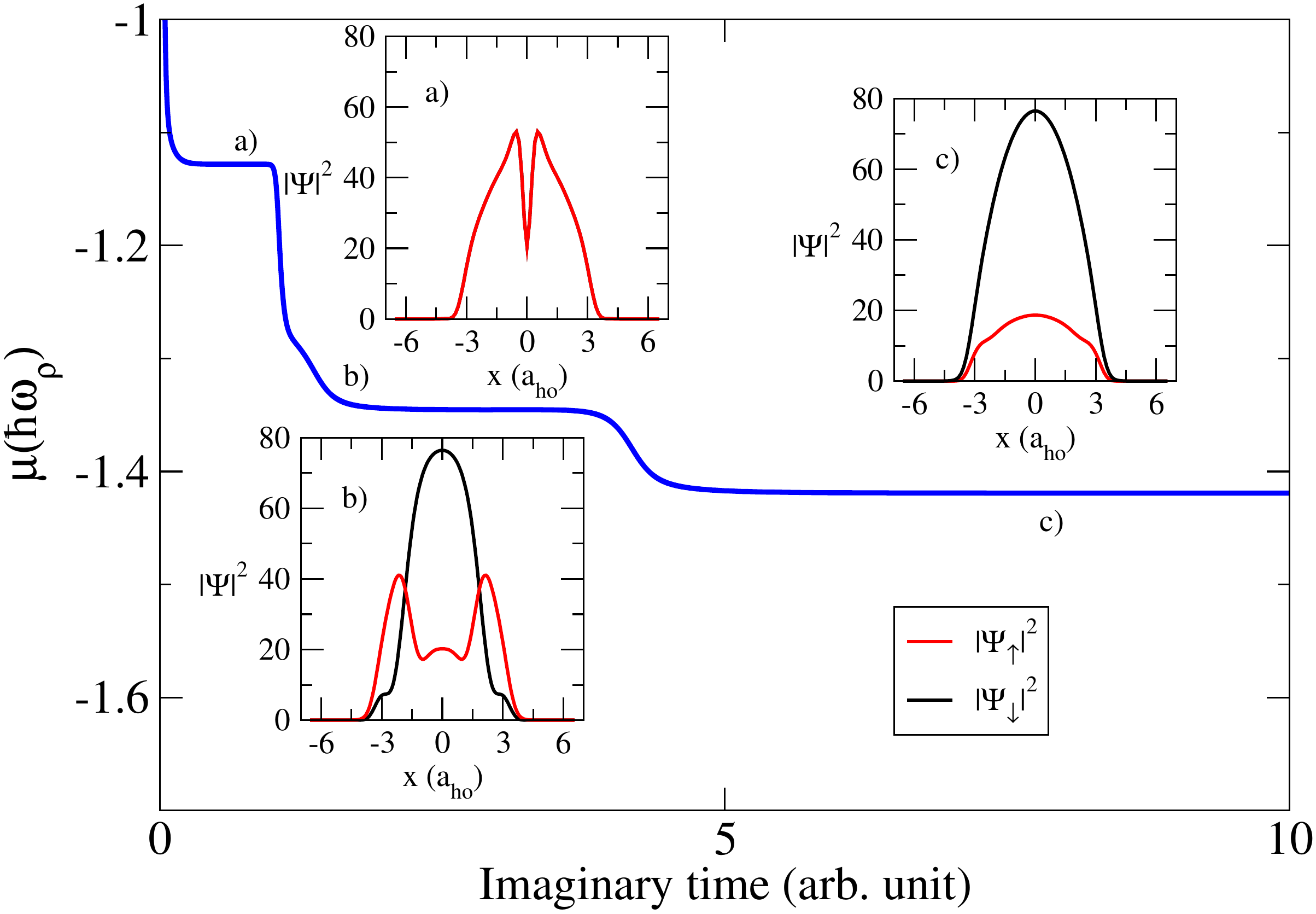}\\
\vspace*{-6.5mm}
\includegraphics[width=0.97\linewidth]{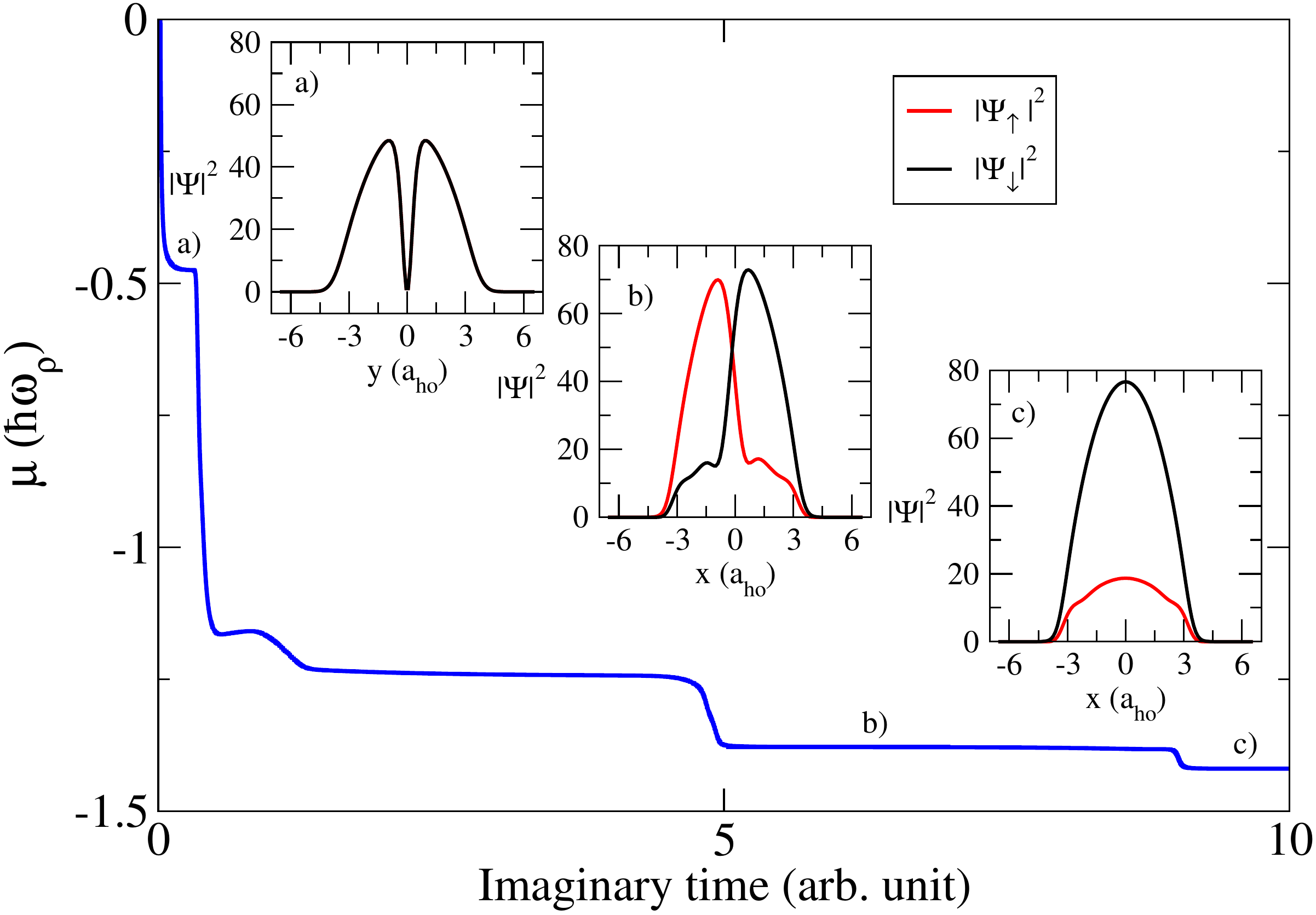}
\caption{Same as Fig. \ref{Fig:imagJV_SM} for JVs in the PW phase  
($\Omega=10 \,\mbox{kHz}$).}
\label{Fig:imagJV_PW}
\end{figure}

As commented previously, the PW phase is characterized by a ground state 
presenting population imbalance. However, the JV states do not share this 
feature. Then, by evolving a JV in imaginary time it is possible to find new, 
intermediate excited states before reaching the ground state.
Figure \ref{Fig:imagJV_PW} represents this evolution for JVs (at insets (a)) 
across $x$ (top panel) and $y$ (bottom panel), having the coherent coupling 
$\Omega=10\,\mbox{kHz}$ and SO number $m\Omega/\hbar k_L^2=0.79$. 
The evolution finds several plateaus that 
correspond to solutions of the time-independent version of Eq. (\ref{tdgpe}), 
and include population imbalance (at insets (b)). At the last stage of the 
evolution, the ground state is obtained (at insets (c)).

\begin{figure}[t!]
\centering
\includegraphics[width=0.97\linewidth]{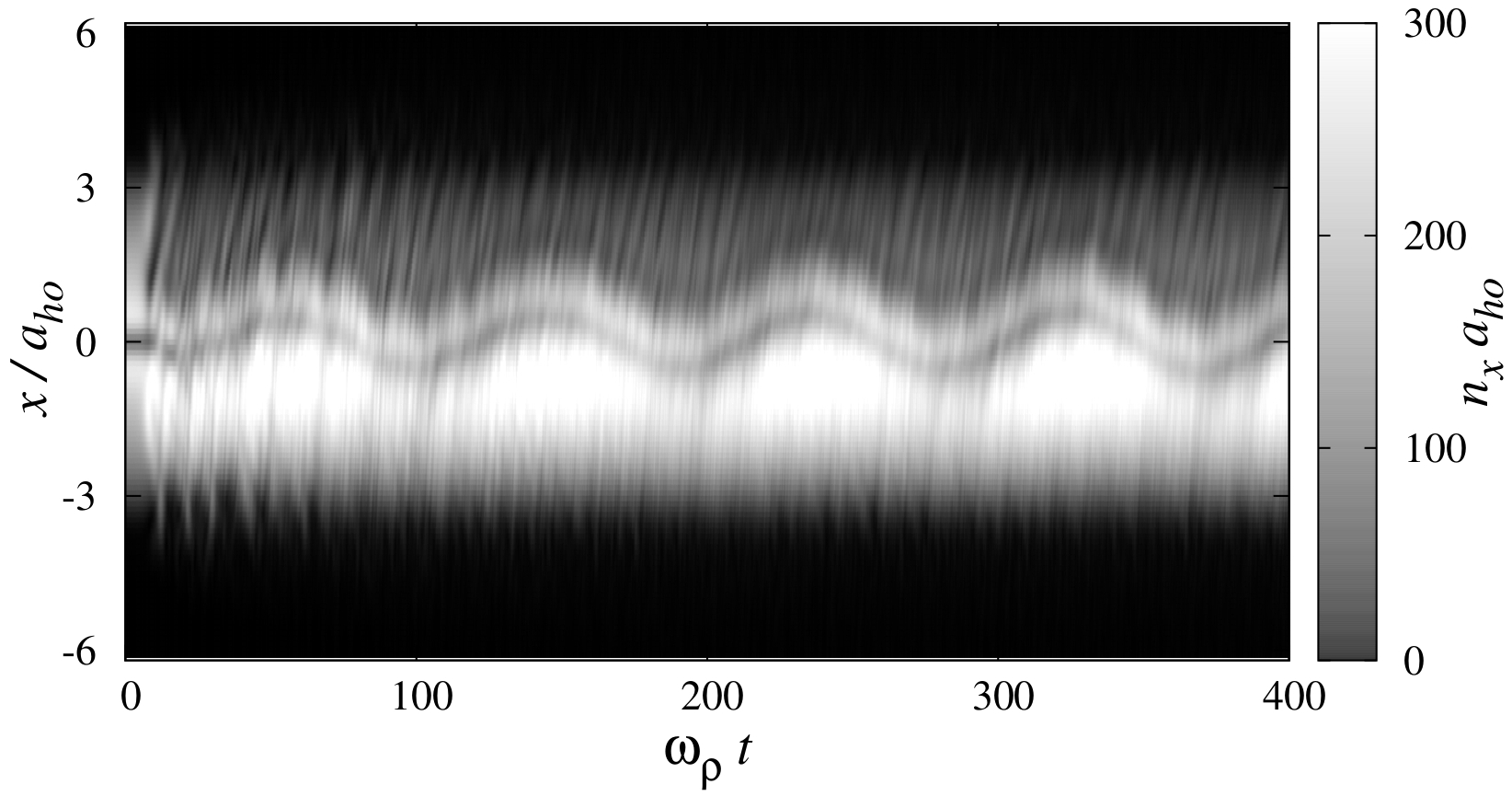}\\
\vspace*{-0.68cm}
\includegraphics[width=0.97\linewidth]{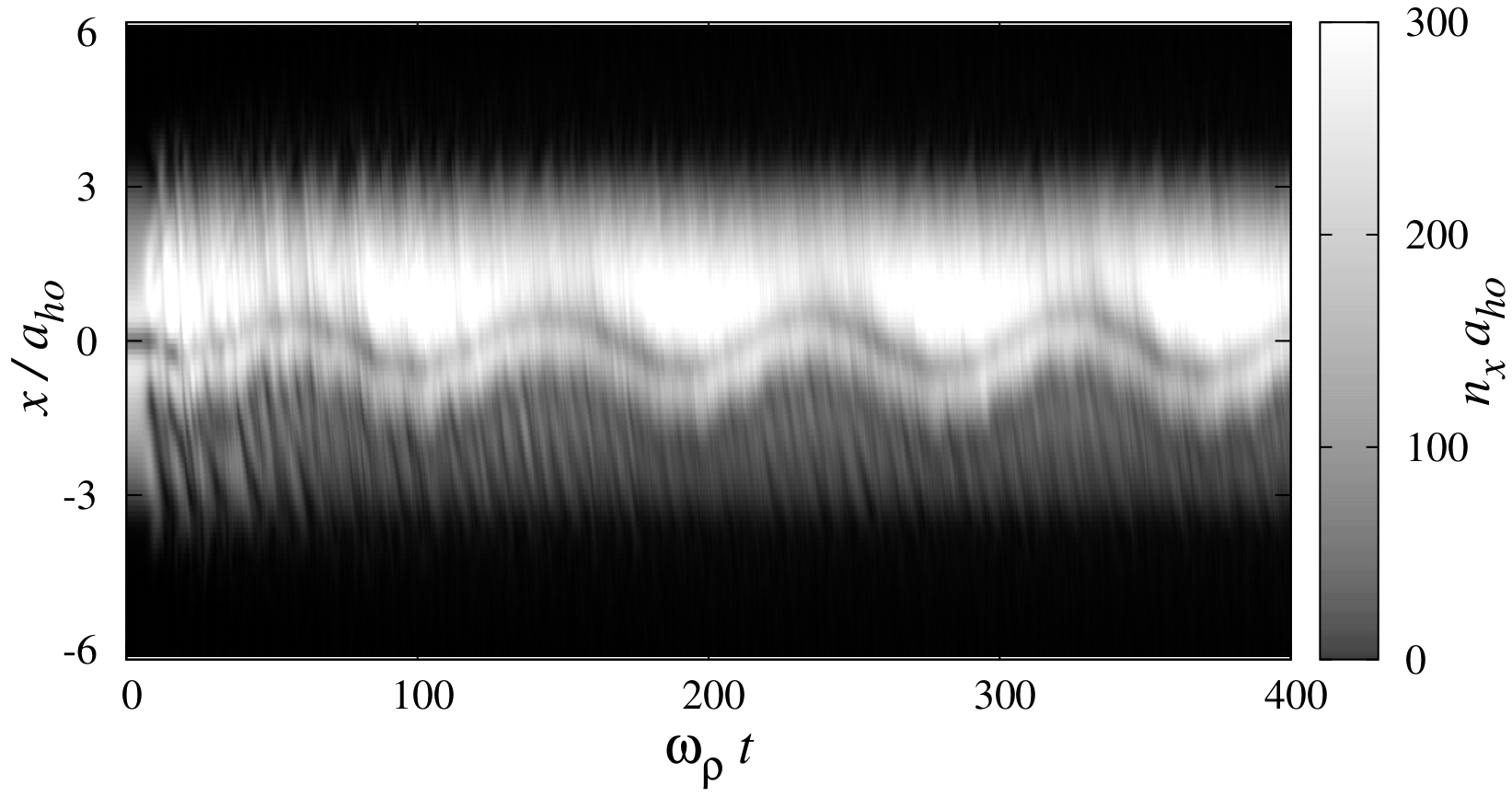}
\caption{Real time evolution of the Josephson vortex, in the $x$-direction, 
shown in the top panel of Fig. \ref{Fig:imagJV_PW}. Top (bottom) panel shows 
the dimensionless, spin-$\uparrow$ (spin-$\downarrow$) density 
after integration over the coordinate $y$.}
\label{Fig:evoJVx_PW}
\end{figure}

In the PW phase, the JV decay is different to the picture outlined before in 
Fig. \ref{Fig:snake}. The characteristic population imbalance of the ground 
state plays a crucial role in the decay dynamics. For the JV across $y$, a big 
population imbalance appears, the density of one of the components is highly 
depleted, and no vortices are finally left over. In contrast, the decay of the 
JV across $x$ is driven by phase separation. 
This phenomenon occurs at the first stage through the formation of 
new solitonic structures during the separation that finally decay into 
vortices. The resulting configuration presents again vortex dipoles that live 
near the surface of each condensate component. The vortex dipoles move 
according to the oscillation of the boundary between components, but their 
orientation does not change in time because of the 
alignment of the dipole velocity and the laser wave vector. A whole evolution 
of this type is represented in Fig. \ref{Fig:evoJVx_PW}, where the integrated 
density profile of both spin components are shown for the JV across the 
$x$-direction of Fig. \ref{Fig:imagJV_PW} (inset (a) of the top panel).

\section{Conclusions}
\label{sec:conclusions}

In this work, we have studied the dynamics of JV states in two-component 
BECs with Rashba-Dresselhaus SO coupling, by using a mean-field model based 
in the Gross-Pitaevskii equation. We have considered the appearance of such 
topological states in a variety of dimensional settings, ranging from 1D 
systems with different interaction strengths, to 3D condensates in 
harmonic traps. Regarding the SO coupling, we have focused on systems belonging 
either to the PW phase, whose ground state presents population 
imbalance, or to the SM phase, with zero population imbalance.

In 1D systems, both in homogeneous and in harmonically trapped settings, we 
have reported on stationary states containing doubly charged, static JVs, 
consisting in a bound state of two JVs. Furthermore, in the trapped case, we 
have characterized families of nonlinear solitonic states connecting with 
analytical solutions of the non-interacting system. 

In multidimensional systems, by numerically solving the full 
GPE for disk-shaped settings, we have found stable JVs in a regime of 
parameters typical of current experiments with $^{87}$Rb atoms.
In addition, we have also studied the snake instability operating on 2D JVs, 
which yields to the appearance of vortex dipoles.
Due to the interaction between the SO wave vector, which breaks the rotational 
symmetry of the system, and the dipole velocities, the subsequent 
evolution sharply changes with respect to that of two-component condensates 
without SO coupling. Our results show how the JV decay leads to different 
scenarios depending, on the one hand, on the orientation of the initial JV, 
and, on the other hand, on the dynamical phase (PW or SM) of 
the system. As a future direction, and beyond the scope of the present work, a 
more detailed analysis concerning the appearance of instability modes, and the 
generation of alternative solitonic states, would be worth of being done.

Finally, we want to emphasize the feasibility of the multidimensional JV states 
presented in this work for their experimental realization, either in the stable 
or in the unstable regime. The latter, in particular, opens up an interesting way 
to study the snake instability in two-component condensates, in the same manner 
as it is currently being the case of DS decay in scalar condensates 
\cite{Ku2014,Ku2015}.
Besides, the decay of JVs in SO coupled condensates provides an excellent 
playground to investigate the rich dynamics of vortex dipoles as has been 
previously done  in scalar BECs \cite{Freilich2010,Navarro2013}.

\acknowledgments

We acknowledge financial support from the Spanish MINECO (FIS2011-28617-C02-01 and 
FIS2014-52285-C2-1-P) and the European Regional development Fund, Generalitat de Catalunya 
Grant No. 2014 SGR 401. A.G. is supported by Spanish MECD fellowship FPU13/02106. 

\bibliography{bibtex}

\begin{thebibliography}{48}%
\makeatletter
\providecommand \@ifxundefined [1]{%
 \@ifx{#1\undefined}
}%
\providecommand \@ifnum [1]{%
 \ifnum #1\expandafter \@firstoftwo
 \else \expandafter \@secondoftwo
 \fi
}%
\providecommand \@ifx [1]{%
 \ifx #1\expandafter \@firstoftwo
 \else \expandafter \@secondoftwo
 \fi
}%
\providecommand \natexlab [1]{#1}%
\providecommand \enquote  [1]{``#1''}%
\providecommand \bibnamefont  [1]{#1}%
\providecommand \bibfnamefont [1]{#1}%
\providecommand \citenamefont [1]{#1}%
\providecommand \href@noop [0]{\@secondoftwo}%
\providecommand \href [0]{\begingroup \@sanitize@url \@href}%
\providecommand \@href[1]{\@@startlink{#1}\@@href}%
\providecommand \@@href[1]{\endgroup#1\@@endlink}%
\providecommand \@sanitize@url [0]{\catcode `\\12\catcode `\$12\catcode
  `\&12\catcode `\#12\catcode `\^12\catcode `\_12\catcode `\%12\relax}%
\providecommand \@@startlink[1]{}%
\providecommand \@@endlink[0]{}%
\providecommand \url  [0]{\begingroup\@sanitize@url \@url }%
\providecommand \@url [1]{\endgroup\@href {#1}{\urlprefix }}%
\providecommand \urlprefix  [0]{URL }%
\providecommand \Eprint [0]{\href }%
\providecommand \doibase [0]{http://dx.doi.org/}%
\providecommand \selectlanguage [0]{\@gobble}%
\providecommand \bibinfo  [0]{\@secondoftwo}%
\providecommand \bibfield  [0]{\@secondoftwo}%
\providecommand \translation [1]{[#1]}%
\providecommand \BibitemOpen [0]{}%
\providecommand \bibitemStop [0]{}%
\providecommand \bibitemNoStop [0]{.\EOS\space}%
\providecommand \EOS [0]{\spacefactor3000\relax}%
\providecommand \BibitemShut  [1]{\csname bibitem#1\endcsname}%
\let\auto@bib@innerbib\@empty
\bibitem [{\citenamefont {Dalibard}\ \emph {et~al.}(2011)\citenamefont
  {Dalibard}, \citenamefont {Gerbier}, \citenamefont {Juzeliunas},\ and\
  \citenamefont {Ohberg}}]{Dalibard2011}%
  \BibitemOpen
  \bibfield  {author} {\bibinfo {author} {\bibfnamefont {J.}~\bibnamefont
  {Dalibard}}, \bibinfo {author} {\bibfnamefont {F.}~\bibnamefont {Gerbier}},
  \bibinfo {author} {\bibfnamefont {G.}~\bibnamefont {Juzeliunas}}, \ and\
  \bibinfo {author} {\bibfnamefont {P.}~\bibnamefont {Ohberg}},\ }\href@noop {}
  {\bibfield  {journal} {\bibinfo  {journal} {Rev. Mod. Phys.}\ }\textbf
  {\bibinfo {volume} {83}},\ \bibinfo {pages} {1523} (\bibinfo {year}
  {2011})}\BibitemShut {NoStop}%
\bibitem [{\citenamefont {Goldman}\ \emph {et~al.}(2014)\citenamefont
  {Goldman}, \citenamefont {Juzeliunas}, \citenamefont {Ohberg},\ and\
  \citenamefont {Spielman}}]{Goldman2014}%
  \BibitemOpen
  \bibfield  {author} {\bibinfo {author} {\bibfnamefont {N.}~\bibnamefont
  {Goldman}}, \bibinfo {author} {\bibfnamefont {G.}~\bibnamefont {Juzeliunas}},
  \bibinfo {author} {\bibfnamefont {P.}~\bibnamefont {Ohberg}}, \ and\ \bibinfo
  {author} {\bibfnamefont {I.~B.}\ \bibnamefont {Spielman}},\ }\href@noop {}
  {\bibfield  {journal} {\bibinfo  {journal} {Rep. Prog. Phys.}\ }\textbf
  {\bibinfo {volume} {77}},\ \bibinfo {pages} {126401} (\bibinfo {year}
  {2014})}\BibitemShut {NoStop}%
\bibitem [{\citenamefont {Stanescu}\ \emph {et~al.}(2008)\citenamefont
  {Stanescu}, \citenamefont {Anderson},\ and\ \citenamefont
  {Galitski}}]{Stanescu2008}%
  \BibitemOpen
  \bibfield  {author} {\bibinfo {author} {\bibfnamefont {T.~D.}\ \bibnamefont
  {Stanescu}}, \bibinfo {author} {\bibfnamefont {B.}~\bibnamefont {Anderson}},
  \ and\ \bibinfo {author} {\bibfnamefont {V.}~\bibnamefont {Galitski}},\
  }\href@noop {} {\bibfield  {journal} {\bibinfo  {journal} {Phys. Rev. A}\
  }\textbf {\bibinfo {volume} {78}},\ \bibinfo {pages} {023616} (\bibinfo
  {year} {2008})}\BibitemShut {NoStop}%
\bibitem [{\citenamefont {Lin}\ \emph {et~al.}(2009)\citenamefont {Lin},
  \citenamefont {Compton}, \citenamefont {{Jim\'enez}-{Garc\'ia}},
  \citenamefont {Porto},\ and\ \citenamefont {Spielman}}]{Lin2009}%
  \BibitemOpen
  \bibfield  {author} {\bibinfo {author} {\bibfnamefont {Y.-J.}\ \bibnamefont
  {Lin}}, \bibinfo {author} {\bibfnamefont {R.~L.}\ \bibnamefont {Compton}},
  \bibinfo {author} {\bibfnamefont {K.}~\bibnamefont {{Jim\'enez}-{Garc\'ia}}},
  \bibinfo {author} {\bibfnamefont {J.~V.}\ \bibnamefont {Porto}}, \ and\
  \bibinfo {author} {\bibfnamefont {I.~B.}\ \bibnamefont {Spielman}},\
  }\href@noop {} {\bibfield  {journal} {\bibinfo  {journal} {Nature}\ }\textbf
  {\bibinfo {volume} {462}},\ \bibinfo {pages} {628} (\bibinfo {year}
  {2009})}\BibitemShut {NoStop}%
\bibitem [{\citenamefont {Lin}\ \emph {et~al.}(2011)\citenamefont {Lin},
  \citenamefont {{Jim\'enez}-{Garc\'ia}},\ and\ \citenamefont
  {Spielman}}]{Lin2011}%
  \BibitemOpen
  \bibfield  {author} {\bibinfo {author} {\bibfnamefont {Y.-J.}\ \bibnamefont
  {Lin}}, \bibinfo {author} {\bibfnamefont {K.}~\bibnamefont
  {{Jim\'enez}-{Garc\'ia}}}, \ and\ \bibinfo {author} {\bibfnamefont {I.~B.}\
  \bibnamefont {Spielman}},\ }\href@noop {} {\bibfield  {journal} {\bibinfo
  {journal} {Nature}\ }\textbf {\bibinfo {volume} {471}},\ \bibinfo {pages}
  {83} (\bibinfo {year} {2011})}\BibitemShut {NoStop}%
\bibitem [{\citenamefont {Galitski}\ and\ \citenamefont
  {Spielman}(2013)}]{Galitski2013}%
  \BibitemOpen
  \bibfield  {author} {\bibinfo {author} {\bibfnamefont {V.}~\bibnamefont
  {Galitski}}\ and\ \bibinfo {author} {\bibfnamefont {I.~B.}\ \bibnamefont
  {Spielman}},\ }\href@noop {} {\bibfield  {journal} {\bibinfo  {journal}
  {Nature}\ }\textbf {\bibinfo {volume} {494}},\ \bibinfo {pages} {49}
  (\bibinfo {year} {2013})}\BibitemShut {NoStop}%
\bibitem [{\citenamefont {Sinha}\ \emph {et~al.}(2011)\citenamefont {Sinha},
  \citenamefont {Nath},\ and\ \citenamefont {Santos}}]{Sinha2011}%
  \BibitemOpen
  \bibfield  {author} {\bibinfo {author} {\bibfnamefont {S.}~\bibnamefont
  {Sinha}}, \bibinfo {author} {\bibfnamefont {R.}~\bibnamefont {Nath}}, \ and\
  \bibinfo {author} {\bibfnamefont {L.}~\bibnamefont {Santos}},\ }\href@noop {}
  {\bibfield  {journal} {\bibinfo  {journal} {Phys. Rev. Lett.}\ }\textbf
  {\bibinfo {volume} {107}},\ \bibinfo {pages} {270401} (\bibinfo {year}
  {2011})}\BibitemShut {NoStop}%
\bibitem [{\citenamefont {Hamner}\ \emph {et~al.}(2015)\citenamefont {Hamner},
  \citenamefont {Zhang}, \citenamefont {Khamehchi}, \citenamefont {Davis},\
  and\ \citenamefont {Engels}}]{Hamner2015}%
  \BibitemOpen
  \bibfield  {author} {\bibinfo {author} {\bibfnamefont {C.}~\bibnamefont
  {Hamner}}, \bibinfo {author} {\bibfnamefont {Y.}~\bibnamefont {Zhang}},
  \bibinfo {author} {\bibfnamefont {M.}~\bibnamefont {Khamehchi}}, \bibinfo
  {author} {\bibfnamefont {M.~J.}\ \bibnamefont {Davis}}, \ and\ \bibinfo
  {author} {\bibfnamefont {P.}~\bibnamefont {Engels}},\ }\href@noop {}
  {\bibfield  {journal} {\bibinfo  {journal} {Phys. Rev. Lett.}\ }\textbf
  {\bibinfo {volume} {114}},\ \bibinfo {pages} {070401} (\bibinfo {year}
  {2015})}\BibitemShut {NoStop}%
\bibitem [{\citenamefont {Zhang}\ and\ \citenamefont
  {Zhang}(2013)}]{Zhang2013}%
  \BibitemOpen
  \bibfield  {author} {\bibinfo {author} {\bibfnamefont {Y.}~\bibnamefont
  {Zhang}}\ and\ \bibinfo {author} {\bibfnamefont {C.}~\bibnamefont {Zhang}},\
  }\href@noop {} {\bibfield  {journal} {\bibinfo  {journal} {Phys. Rev. A}\
  }\textbf {\bibinfo {volume} {87}},\ \bibinfo {pages} {023611} (\bibinfo
  {year} {2013})}\BibitemShut {NoStop}%
\bibitem [{\citenamefont {Chen}\ \emph {et~al.}(2014)\citenamefont {Chen},
  \citenamefont {Rabinovic}, \citenamefont {Anderson},\ and\ \citenamefont
  {Santos}}]{Chen2014}%
  \BibitemOpen
  \bibfield  {author} {\bibinfo {author} {\bibfnamefont {X.}~\bibnamefont
  {Chen}}, \bibinfo {author} {\bibfnamefont {M.}~\bibnamefont {Rabinovic}},
  \bibinfo {author} {\bibfnamefont {B.~M.}\ \bibnamefont {Anderson}}, \ and\
  \bibinfo {author} {\bibfnamefont {L.}~\bibnamefont {Santos}},\ }\href@noop {}
  {\bibfield  {journal} {\bibinfo  {journal} {Phys. Rev. A}\ }\textbf {\bibinfo
  {volume} {90}},\ \bibinfo {pages} {043632} (\bibinfo {year}
  {2014})}\BibitemShut {NoStop}%
\bibitem [{\citenamefont {Mivehvar}\ and\ \citenamefont
  {Feder}(2015)}]{Mivehvar2015}%
  \BibitemOpen
  \bibfield  {author} {\bibinfo {author} {\bibfnamefont {F.}~\bibnamefont
  {Mivehvar}}\ and\ \bibinfo {author} {\bibfnamefont {D.~L.}\ \bibnamefont
  {Feder}},\ }\href@noop {} {\bibfield  {journal} {\bibinfo  {journal} {Phys.
  Rev. A}\ }\textbf {\bibinfo {volume} {92}},\ \bibinfo {pages} {023611}
  (\bibinfo {year} {2015})}\BibitemShut {NoStop}%
\bibitem [{\citenamefont {Radic}\ \emph {et~al.}(2011)\citenamefont {Radic},
  \citenamefont {Sedrakyan}, \citenamefont {Spielman},\ and\ \citenamefont
  {Galitski}}]{Radic2011}%
  \BibitemOpen
  \bibfield  {author} {\bibinfo {author} {\bibfnamefont {J.}~\bibnamefont
  {Radic}}, \bibinfo {author} {\bibfnamefont {T.~A.}\ \bibnamefont
  {Sedrakyan}}, \bibinfo {author} {\bibfnamefont {I.~B.}\ \bibnamefont
  {Spielman}}, \ and\ \bibinfo {author} {\bibfnamefont {V.}~\bibnamefont
  {Galitski}},\ }\href@noop {} {\bibfield  {journal} {\bibinfo  {journal}
  {Phys. Rev. A}\ }\textbf {\bibinfo {volume} {84}},\ \bibinfo {pages} {063624}
  (\bibinfo {year} {2011})}\BibitemShut {NoStop}%
\bibitem [{\citenamefont {Ramachandhran}\ \emph {et~al.}(2012)\citenamefont
  {Ramachandhran}, \citenamefont {Opanchuk}, \citenamefont {Liu}, \citenamefont
  {Pu}, \citenamefont {Drummond},\ and\ \citenamefont
  {Hu}}]{Ramachandhran2012}%
  \BibitemOpen
  \bibfield  {author} {\bibinfo {author} {\bibfnamefont {B.}~\bibnamefont
  {Ramachandhran}}, \bibinfo {author} {\bibfnamefont {B.}~\bibnamefont
  {Opanchuk}}, \bibinfo {author} {\bibfnamefont {X.-J.}\ \bibnamefont {Liu}},
  \bibinfo {author} {\bibfnamefont {H.}~\bibnamefont {Pu}}, \bibinfo {author}
  {\bibfnamefont {P.~D.}\ \bibnamefont {Drummond}}, \ and\ \bibinfo {author}
  {\bibfnamefont {H.}~\bibnamefont {Hu}},\ }\href@noop {} {\bibfield  {journal}
  {\bibinfo  {journal} {Phys. Rev. A}\ }\textbf {\bibinfo {volume} {85}},\
  \bibinfo {pages} {023606} (\bibinfo {year} {2012})}\BibitemShut {NoStop}%
\bibitem [{\citenamefont {Martone}\ \emph {et~al.}(2012)\citenamefont
  {Martone}, \citenamefont {Li}, \citenamefont {Pitaevskii},\ and\
  \citenamefont {Stringari}}]{Martone2012}%
  \BibitemOpen
  \bibfield  {author} {\bibinfo {author} {\bibfnamefont {G.~I.}\ \bibnamefont
  {Martone}}, \bibinfo {author} {\bibfnamefont {Y.}~\bibnamefont {Li}},
  \bibinfo {author} {\bibfnamefont {L.~P.}\ \bibnamefont {Pitaevskii}}, \ and\
  \bibinfo {author} {\bibfnamefont {S.}~\bibnamefont {Stringari}},\ }\href@noop
  {} {\bibfield  {journal} {\bibinfo  {journal} {Phys. Rev. A}\ }\textbf
  {\bibinfo {volume} {86}},\ \bibinfo {pages} {063621} (\bibinfo {year}
  {2012})}\BibitemShut {NoStop}%
\bibitem [{\citenamefont {Zhang}\ \emph {et~al.}(2012)\citenamefont {Zhang},
  \citenamefont {Fu}, \citenamefont {Wang},\ and\ \citenamefont
  {Zhu}}]{Zhang2012b}%
  \BibitemOpen
  \bibfield  {author} {\bibinfo {author} {\bibfnamefont {D.-W.}\ \bibnamefont
  {Zhang}}, \bibinfo {author} {\bibfnamefont {L.-B.}\ \bibnamefont {Fu}},
  \bibinfo {author} {\bibfnamefont {Z.~D.}\ \bibnamefont {Wang}}, \ and\
  \bibinfo {author} {\bibfnamefont {S.-L.}\ \bibnamefont {Zhu}},\ }\href@noop
  {} {\bibfield  {journal} {\bibinfo  {journal} {Phys. Rev. A}\ }\textbf
  {\bibinfo {volume} {85}},\ \bibinfo {pages} {043609} (\bibinfo {year}
  {2012})}\BibitemShut {NoStop}%
\bibitem [{\citenamefont {{Garc\'ia}-March}\ \emph {et~al.}(2014)\citenamefont
  {{Garc\'ia}-March}, \citenamefont {Mazzarella}, \citenamefont {Dell'Anna},
  \citenamefont {{Juli\'a-D\'iaz}}, \citenamefont {Salasnich},\ and\
  \citenamefont {Polls}}]{GarciaMarch2014}%
  \BibitemOpen
  \bibfield  {author} {\bibinfo {author} {\bibfnamefont {M.~A.}\ \bibnamefont
  {{Garc\'ia}-March}}, \bibinfo {author} {\bibfnamefont {G.}~\bibnamefont
  {Mazzarella}}, \bibinfo {author} {\bibfnamefont {L.}~\bibnamefont
  {Dell'Anna}}, \bibinfo {author} {\bibfnamefont {B.}~\bibnamefont
  {{Juli\'a-D\'iaz}}}, \bibinfo {author} {\bibfnamefont {L.}~\bibnamefont
  {Salasnich}}, \ and\ \bibinfo {author} {\bibfnamefont {A.}~\bibnamefont
  {Polls}},\ }\href@noop {} {\bibfield  {journal} {\bibinfo  {journal} {Phys.
  Rev. A}\ }\textbf {\bibinfo {volume} {89}},\ \bibinfo {pages} {063607}
  (\bibinfo {year} {2014})}\BibitemShut {NoStop}%
\bibitem [{\citenamefont {Li}\ \emph {et~al.}(2012{\natexlab{a}})\citenamefont
  {Li}, \citenamefont {Pitaevskii},\ and\ \citenamefont {Stringari}}]{Li2012a}%
  \BibitemOpen
  \bibfield  {author} {\bibinfo {author} {\bibfnamefont {Y.}~\bibnamefont
  {Li}}, \bibinfo {author} {\bibfnamefont {L.~P.}\ \bibnamefont {Pitaevskii}},
  \ and\ \bibinfo {author} {\bibfnamefont {S.}~\bibnamefont {Stringari}},\
  }\href@noop {} {\bibfield  {journal} {\bibinfo  {journal} {Phys. Rev. Lett.}\
  }\textbf {\bibinfo {volume} {108}},\ \bibinfo {pages} {225301} (\bibinfo
  {year} {2012}{\natexlab{a}})}\BibitemShut {NoStop}%
\bibitem [{\citenamefont {Li}\ \emph {et~al.}(2012{\natexlab{b}})\citenamefont
  {Li}, \citenamefont {Martone},\ and\ \citenamefont {Stringari}}]{Li2012b}%
  \BibitemOpen
  \bibfield  {author} {\bibinfo {author} {\bibfnamefont {Y.}~\bibnamefont
  {Li}}, \bibinfo {author} {\bibfnamefont {G.~I.}\ \bibnamefont {Martone}}, \
  and\ \bibinfo {author} {\bibfnamefont {S.}~\bibnamefont {Stringari}},\
  }\href@noop {} {\bibfield  {journal} {\bibinfo  {journal} {Europhys. Lett.}\
  }\textbf {\bibinfo {volume} {99}},\ \bibinfo {pages} {56008} (\bibinfo {year}
  {2012}{\natexlab{b}})}\BibitemShut {NoStop}%
\bibitem [{\citenamefont {Li}\ \emph {et~al.}(2015)\citenamefont {Li},
  \citenamefont {Martone},\ and\ \citenamefont {Stringari}}]{Li2015}%
  \BibitemOpen
  \bibfield  {author} {\bibinfo {author} {\bibfnamefont {Y.}~\bibnamefont
  {Li}}, \bibinfo {author} {\bibfnamefont {G.~I.}\ \bibnamefont {Martone}}, \
  and\ \bibinfo {author} {\bibfnamefont {S.}~\bibnamefont {Stringari}},\
  }\enquote {\bibinfo {title} {Annual review of cold atoms and molecules},}\ \
  (\bibinfo {year} {2015})\ Chap.\ \bibinfo {chapter} {SPIN-ORBIT-COUPLED
  BOSE-EINSTEIN CONDENSATES}, p.\ \bibinfo {pages} {201}\BibitemShut {NoStop}%
\bibitem [{\citenamefont {Williams}\ and\ \citenamefont
  {Holland}(1999)}]{Williams1999a}%
  \BibitemOpen
  \bibfield  {author} {\bibinfo {author} {\bibfnamefont {J.~E.}\ \bibnamefont
  {Williams}}\ and\ \bibinfo {author} {\bibfnamefont {M.~J.}\ \bibnamefont
  {Holland}},\ }\href@noop {} {\bibfield  {journal} {\bibinfo  {journal}
  {Nature}\ }\textbf {\bibinfo {volume} {401}},\ \bibinfo {pages} {568}
  (\bibinfo {year} {1999})}\BibitemShut {NoStop}%
\bibitem [{\citenamefont {Williams}\ \emph {et~al.}(1999)\citenamefont
  {Williams}, \citenamefont {Walser}, \citenamefont {Cooper}, \citenamefont
  {Cornell},\ and\ \citenamefont {Holland}}]{Williams1999b}%
  \BibitemOpen
  \bibfield  {author} {\bibinfo {author} {\bibfnamefont {J.~E.}\ \bibnamefont
  {Williams}}, \bibinfo {author} {\bibfnamefont {R.}~\bibnamefont {Walser}},
  \bibinfo {author} {\bibfnamefont {J.}~\bibnamefont {Cooper}}, \bibinfo
  {author} {\bibfnamefont {E.~A.}\ \bibnamefont {Cornell}}, \ and\ \bibinfo
  {author} {\bibfnamefont {M.~J.}\ \bibnamefont {Holland}},\ }\href@noop {}
  {\bibfield  {journal} {\bibinfo  {journal} {Phys. Rev. A}\ }\textbf {\bibinfo
  {volume} {59}},\ \bibinfo {pages} {R31(R)} (\bibinfo {year}
  {1999})}\BibitemShut {NoStop}%
\bibitem [{\citenamefont {Gross}\ \emph {et~al.}(2010)\citenamefont {Gross},
  \citenamefont {Zibold}, \citenamefont {Nicklas}, \citenamefont {{Est\`eve}},\
  and\ \citenamefont {Oberthaler}}]{Gross2010}%
  \BibitemOpen
  \bibfield  {author} {\bibinfo {author} {\bibfnamefont {C.}~\bibnamefont
  {Gross}}, \bibinfo {author} {\bibfnamefont {T.}~\bibnamefont {Zibold}},
  \bibinfo {author} {\bibfnamefont {E.}~\bibnamefont {Nicklas}}, \bibinfo
  {author} {\bibfnamefont {J.}~\bibnamefont {{Est\`eve}}}, \ and\ \bibinfo
  {author} {\bibfnamefont {M.~K.}\ \bibnamefont {Oberthaler}},\ }\href@noop {}
  {\bibfield  {journal} {\bibinfo  {journal} {Nature}\ }\textbf {\bibinfo
  {volume} {464}},\ \bibinfo {pages} {1165} (\bibinfo {year}
  {2010})}\BibitemShut {NoStop}%
\bibitem [{\citenamefont {Smerzi}\ \emph {et~al.}(1997)\citenamefont {Smerzi},
  \citenamefont {Fantoni}, \citenamefont {Giovanazzi},\ and\ \citenamefont
  {Shenoy}}]{Smerzi1997}%
  \BibitemOpen
  \bibfield  {author} {\bibinfo {author} {\bibfnamefont {A.}~\bibnamefont
  {Smerzi}}, \bibinfo {author} {\bibfnamefont {S.}~\bibnamefont {Fantoni}},
  \bibinfo {author} {\bibfnamefont {S.}~\bibnamefont {Giovanazzi}}, \ and\
  \bibinfo {author} {\bibfnamefont {S.~R.}\ \bibnamefont {Shenoy}},\
  }\href@noop {} {\bibfield  {journal} {\bibinfo  {journal} {Phys. Rev. Lett.}\
  }\textbf {\bibinfo {volume} {79}},\ \bibinfo {pages} {4950} (\bibinfo {year}
  {1997})}\BibitemShut {NoStop}%
\bibitem [{\citenamefont {Raghavan}\ \emph {et~al.}(1999)\citenamefont
  {Raghavan}, \citenamefont {Smerzi}, \citenamefont {Fantoni},\ and\
  \citenamefont {Shenoy}}]{Raghavan1999}%
  \BibitemOpen
  \bibfield  {author} {\bibinfo {author} {\bibfnamefont {S.}~\bibnamefont
  {Raghavan}}, \bibinfo {author} {\bibfnamefont {A.}~\bibnamefont {Smerzi}},
  \bibinfo {author} {\bibfnamefont {S.}~\bibnamefont {Fantoni}}, \ and\
  \bibinfo {author} {\bibfnamefont {S.~R.}\ \bibnamefont {Shenoy}},\
  }\href@noop {} {\bibfield  {journal} {\bibinfo  {journal} {Phys. Rev. A}\
  }\textbf {\bibinfo {volume} {59}},\ \bibinfo {pages} {620} (\bibinfo {year}
  {1999})}\BibitemShut {NoStop}%
\bibitem [{\citenamefont {Albiez}\ \emph {et~al.}(2005)\citenamefont {Albiez},
  \citenamefont {Gati}, \citenamefont {{F\"olling}}, \citenamefont {Hunsmann},
  \citenamefont {Cristiani},\ and\ \citenamefont {Oberthaler}}]{Albiez2005}%
  \BibitemOpen
  \bibfield  {author} {\bibinfo {author} {\bibfnamefont {M.}~\bibnamefont
  {Albiez}}, \bibinfo {author} {\bibfnamefont {R.}~\bibnamefont {Gati}},
  \bibinfo {author} {\bibfnamefont {J.}~\bibnamefont {{F\"olling}}}, \bibinfo
  {author} {\bibfnamefont {S.}~\bibnamefont {Hunsmann}}, \bibinfo {author}
  {\bibfnamefont {M.}~\bibnamefont {Cristiani}}, \ and\ \bibinfo {author}
  {\bibfnamefont {M.~K.}\ \bibnamefont {Oberthaler}},\ }\href@noop {}
  {\bibfield  {journal} {\bibinfo  {journal} {Phys. Rev. Lett.}\ }\textbf
  {\bibinfo {volume} {95}},\ \bibinfo {pages} {010402} (\bibinfo {year}
  {2005})}\BibitemShut {NoStop}%
\bibitem [{\citenamefont {Levy}\ \emph {et~al.}(2007)\citenamefont {Levy},
  \citenamefont {Lahoud}, \citenamefont {Shomroni},\ and\ \citenamefont
  {Steinhauer}}]{Levy2007}%
  \BibitemOpen
  \bibfield  {author} {\bibinfo {author} {\bibfnamefont {S.}~\bibnamefont
  {Levy}}, \bibinfo {author} {\bibfnamefont {E.}~\bibnamefont {Lahoud}},
  \bibinfo {author} {\bibfnamefont {I.}~\bibnamefont {Shomroni}}, \ and\
  \bibinfo {author} {\bibfnamefont {J.}~\bibnamefont {Steinhauer}},\
  }\href@noop {} {\bibfield  {journal} {\bibinfo  {journal} {Nature}\ }\textbf
  {\bibinfo {volume} {449}},\ \bibinfo {pages} {579} (\bibinfo {year}
  {2007})}\BibitemShut {NoStop}%
\bibitem [{\citenamefont {Son}\ and\ \citenamefont
  {Stephanov}(2002)}]{Son2002}%
  \BibitemOpen
  \bibfield  {author} {\bibinfo {author} {\bibfnamefont {D.~T.}\ \bibnamefont
  {Son}}\ and\ \bibinfo {author} {\bibfnamefont {M.~A.}\ \bibnamefont
  {Stephanov}},\ }\href@noop {} {\bibfield  {journal} {\bibinfo  {journal}
  {Phys. Rev. A}\ }\textbf {\bibinfo {volume} {65}},\ \bibinfo {pages} {063621}
  (\bibinfo {year} {2002})}\BibitemShut {NoStop}%
\bibitem [{\citenamefont {Kaurov}\ and\ \citenamefont
  {Kuklov}(2005)}]{Kaurov2005}%
  \BibitemOpen
  \bibfield  {author} {\bibinfo {author} {\bibfnamefont {V.~M.}\ \bibnamefont
  {Kaurov}}\ and\ \bibinfo {author} {\bibfnamefont {A.~B.}\ \bibnamefont
  {Kuklov}},\ }\href@noop {} {\bibfield  {journal} {\bibinfo  {journal} {Phys.
  Rev. A}\ }\textbf {\bibinfo {volume} {71}},\ \bibinfo {pages} {011601(R)}
  (\bibinfo {year} {2005})}\BibitemShut {NoStop}%
\bibitem [{\citenamefont {Kaurov}\ and\ \citenamefont
  {Kuklov}(2006)}]{Kaurov2006}%
  \BibitemOpen
  \bibfield  {author} {\bibinfo {author} {\bibfnamefont {V.~M.}\ \bibnamefont
  {Kaurov}}\ and\ \bibinfo {author} {\bibfnamefont {A.~B.}\ \bibnamefont
  {Kuklov}},\ }\href@noop {} {\bibfield  {journal} {\bibinfo  {journal} {Phys.
  Rev. A}\ }\textbf {\bibinfo {volume} {73}},\ \bibinfo {pages} {013627}
  (\bibinfo {year} {2006})}\BibitemShut {NoStop}%
\bibitem [{\citenamefont {Barone}\ and\ \citenamefont
  {Paterno}(1982)}]{Barone1982}%
  \BibitemOpen
  \bibfield  {author} {\bibinfo {author} {\bibfnamefont {A.}~\bibnamefont
  {Barone}}\ and\ \bibinfo {author} {\bibfnamefont {G.}~\bibnamefont
  {Paterno}},\ }\href@noop {} {\emph {\bibinfo {title} {Physics and
  Applications of the Josephson Effect}}}\ (\bibinfo  {publisher} {Wiley and
  Sons Inc.},\ \bibinfo {year} {1982})\BibitemShut {NoStop}%
\bibitem [{\citenamefont {Su}\ \emph {et~al.}(2013)\citenamefont {Su},
  \citenamefont {Gou}, \citenamefont {Bradley}, \citenamefont {Fialko},\ and\
  \citenamefont {Brand}}]{Su2013}%
  \BibitemOpen
  \bibfield  {author} {\bibinfo {author} {\bibfnamefont {S.-W.}\ \bibnamefont
  {Su}}, \bibinfo {author} {\bibfnamefont {S.-C.}\ \bibnamefont {Gou}},
  \bibinfo {author} {\bibfnamefont {A.}~\bibnamefont {Bradley}}, \bibinfo
  {author} {\bibfnamefont {O.}~\bibnamefont {Fialko}}, \ and\ \bibinfo {author}
  {\bibfnamefont {J.}~\bibnamefont {Brand}},\ }\href@noop {} {\bibfield
  {journal} {\bibinfo  {journal} {Phys. Rev. Lett.}\ }\textbf {\bibinfo
  {volume} {110}},\ \bibinfo {pages} {215302} (\bibinfo {year}
  {2013})}\BibitemShut {NoStop}%
\bibitem [{\citenamefont {Achilleos}\ \emph {et~al.}(2013)\citenamefont
  {Achilleos}, \citenamefont {Stockhofe}, \citenamefont {Kevrekidis},
  \citenamefont {Frantzeskakis},\ and\ \citenamefont
  {Schmelcher}}]{Achilleos2013}%
  \BibitemOpen
  \bibfield  {author} {\bibinfo {author} {\bibfnamefont {V.}~\bibnamefont
  {Achilleos}}, \bibinfo {author} {\bibfnamefont {J.}~\bibnamefont
  {Stockhofe}}, \bibinfo {author} {\bibfnamefont {P.~G.}\ \bibnamefont
  {Kevrekidis}}, \bibinfo {author} {\bibfnamefont {D.~J.}\ \bibnamefont
  {Frantzeskakis}}, \ and\ \bibinfo {author} {\bibfnamefont {P.}~\bibnamefont
  {Schmelcher}},\ }\href@noop {} {\bibfield  {journal} {\bibinfo  {journal}
  {Europhys. Lett.}\ }\textbf {\bibinfo {volume} {103}},\ \bibinfo {pages}
  {20002} (\bibinfo {year} {2013})}\BibitemShut {NoStop}%
\bibitem [{\citenamefont {Cao}\ \emph {et~al.}(2015)\citenamefont {Cao},
  \citenamefont {Shan}, \citenamefont {Zhang}, \citenamefont {Qin},\ and\
  \citenamefont {Xu}}]{Cao2015}%
  \BibitemOpen
  \bibfield  {author} {\bibinfo {author} {\bibfnamefont {S.}~\bibnamefont
  {Cao}}, \bibinfo {author} {\bibfnamefont {C.-J.}\ \bibnamefont {Shan}},
  \bibinfo {author} {\bibfnamefont {D.-W.}\ \bibnamefont {Zhang}}, \bibinfo
  {author} {\bibfnamefont {X.}~\bibnamefont {Qin}}, \ and\ \bibinfo {author}
  {\bibfnamefont {J.}~\bibnamefont {Xu}},\ }\href@noop {} {\bibfield  {journal}
  {\bibinfo  {journal} {J. Opt. Soc. Am. B}\ }\textbf {\bibinfo {volume}
  {32}},\ \bibinfo {pages} {201} (\bibinfo {year} {2015})}\BibitemShut
  {NoStop}%
\bibitem [{\citenamefont {Brand}\ and\ \citenamefont
  {Reinhardt}(2001)}]{Brand2001}%
  \BibitemOpen
  \bibfield  {author} {\bibinfo {author} {\bibfnamefont {J.}~\bibnamefont
  {Brand}}\ and\ \bibinfo {author} {\bibfnamefont {W.~P.}\ \bibnamefont
  {Reinhardt}},\ }\href@noop {} {\bibfield  {journal} {\bibinfo  {journal} {J.
  Phys. B: At. Mol. Opt. Phys.}\ }\textbf {\bibinfo {volume} {34}},\ \bibinfo
  {pages} {4} (\bibinfo {year} {2001})}\BibitemShut {NoStop}%
\bibitem [{\citenamefont {{Mu\~noz Mateo}}\ and\ \citenamefont
  {Brand}(2014)}]{MunozMateo2014}%
  \BibitemOpen
  \bibfield  {author} {\bibinfo {author} {\bibfnamefont {A.}~\bibnamefont
  {{Mu\~noz Mateo}}}\ and\ \bibinfo {author} {\bibfnamefont {J.}~\bibnamefont
  {Brand}},\ }\href@noop {} {\bibfield  {journal} {\bibinfo  {journal} {Phys.
  Rev. Lett.}\ }\textbf {\bibinfo {volume} {113}},\ \bibinfo {pages} {255302}
  (\bibinfo {year} {2014})}\BibitemShut {NoStop}%
\bibitem [{\citenamefont {Abad}\ and\ \citenamefont {Recati}(2013)}]{Abad2013}%
  \BibitemOpen
  \bibfield  {author} {\bibinfo {author} {\bibfnamefont {M.}~\bibnamefont
  {Abad}}\ and\ \bibinfo {author} {\bibfnamefont {A.}~\bibnamefont {Recati}},\
  }\href@noop {} {\bibfield  {journal} {\bibinfo  {journal} {Eur. Phys. J. D}\
  }\textbf {\bibinfo {volume} {67}},\ \bibinfo {pages} {148} (\bibinfo {year}
  {2013})}\BibitemShut {NoStop}%
\bibitem [{\citenamefont {Ho}\ and\ \citenamefont {Zhang}(2011)}]{Ho2011}%
  \BibitemOpen
  \bibfield  {author} {\bibinfo {author} {\bibfnamefont {T.-L.}\ \bibnamefont
  {Ho}}\ and\ \bibinfo {author} {\bibfnamefont {S.}~\bibnamefont {Zhang}},\
  }\href@noop {} {\bibfield  {journal} {\bibinfo  {journal} {Phys. Rev. Lett.}\
  }\textbf {\bibinfo {volume} {107}},\ \bibinfo {pages} {150403} (\bibinfo
  {year} {2011})}\BibitemShut {NoStop}%
\bibitem [{\citenamefont {Qadir}\ \emph {et~al.}(2012)\citenamefont {Qadir},
  \citenamefont {Susanto},\ and\ \citenamefont {Matthews}}]{Qadir2012}%
  \BibitemOpen
  \bibfield  {author} {\bibinfo {author} {\bibfnamefont {M.~I.}\ \bibnamefont
  {Qadir}}, \bibinfo {author} {\bibfnamefont {H.}~\bibnamefont {Susanto}}, \
  and\ \bibinfo {author} {\bibfnamefont {P.~C.}\ \bibnamefont {Matthews}},\
  }\href@noop {} {\bibfield  {journal} {\bibinfo  {journal} {J. Phys. B: At.
  Mol. Opt. Phys.}\ }\textbf {\bibinfo {volume} {45}},\ \bibinfo {pages}
  {035004} (\bibinfo {year} {2012})}\BibitemShut {NoStop}%
\bibitem [{\citenamefont {{Gallem\'i}}\ \emph {et~al.}(2015)\citenamefont
  {{Gallem\'i}}, \citenamefont {{Mu\~noz Mateo}}, \citenamefont {Mayol},\ and\
  \citenamefont {Guilleumas}}]{Gallemi2015b}%
  \BibitemOpen
  \bibfield  {author} {\bibinfo {author} {\bibfnamefont {A.}~\bibnamefont
  {{Gallem\'i}}}, \bibinfo {author} {\bibfnamefont {A.}~\bibnamefont {{Mu\~noz
  Mateo}}}, \bibinfo {author} {\bibfnamefont {R.}~\bibnamefont {Mayol}}, \ and\
  \bibinfo {author} {\bibfnamefont {M.}~\bibnamefont {Guilleumas}},\
  }\href@noop {} {\bibfield  {journal} {\bibinfo  {journal} {arXiv:1509.04418}\
  } (\bibinfo {year} {2015})}\BibitemShut {NoStop}%
\bibitem [{\citenamefont {Su}\ \emph {et~al.}(2015)\citenamefont {Su},
  \citenamefont {Gou}, \citenamefont {Liu}, \citenamefont {Bradley},
  \citenamefont {Fialko},\ and\ \citenamefont {Brand}}]{Su2015}%
  \BibitemOpen
  \bibfield  {author} {\bibinfo {author} {\bibfnamefont {S.-W.}\ \bibnamefont
  {Su}}, \bibinfo {author} {\bibfnamefont {S.-C.}\ \bibnamefont {Gou}},
  \bibinfo {author} {\bibfnamefont {I.-K.}\ \bibnamefont {Liu}}, \bibinfo
  {author} {\bibfnamefont {A.~S.}\ \bibnamefont {Bradley}}, \bibinfo {author}
  {\bibfnamefont {O.}~\bibnamefont {Fialko}}, \ and\ \bibinfo {author}
  {\bibfnamefont {J.}~\bibnamefont {Brand}},\ }\href@noop {} {\bibfield
  {journal} {\bibinfo  {journal} {Phys. Rev. A}\ }\textbf {\bibinfo {volume}
  {91}},\ \bibinfo {pages} {023631} (\bibinfo {year} {2015})}\BibitemShut
  {NoStop}%
\bibitem [{\citenamefont {Roditchev}\ \emph {et~al.}(2015)\citenamefont
  {Roditchev}, \citenamefont {Brun}, \citenamefont {Serrier-Garcia},
  \citenamefont {Cuevas}, \citenamefont {Bessa}, \citenamefont {Milosevic},
  \citenamefont {Debontridder}, \citenamefont {Stolyarov},\ and\ \citenamefont
  {Cren}}]{Roditchev2015}%
  \BibitemOpen
  \bibfield  {author} {\bibinfo {author} {\bibfnamefont {D.}~\bibnamefont
  {Roditchev}}, \bibinfo {author} {\bibfnamefont {C.}~\bibnamefont {Brun}},
  \bibinfo {author} {\bibfnamefont {L.}~\bibnamefont {Serrier-Garcia}},
  \bibinfo {author} {\bibfnamefont {J.~C.}\ \bibnamefont {Cuevas}}, \bibinfo
  {author} {\bibfnamefont {V.~H.~L.}\ \bibnamefont {Bessa}}, \bibinfo {author}
  {\bibfnamefont {M.~V.}\ \bibnamefont {Milosevic}}, \bibinfo {author}
  {\bibfnamefont {F.}~\bibnamefont {Debontridder}}, \bibinfo {author}
  {\bibfnamefont {V.}~\bibnamefont {Stolyarov}}, \ and\ \bibinfo {author}
  {\bibfnamefont {T.}~\bibnamefont {Cren}},\ }\href@noop {} {\bibfield
  {journal} {\bibinfo  {journal} {Nat. Phys.}\ }\textbf {\bibinfo {volume}
  {11}},\ \bibinfo {pages} {332} (\bibinfo {year} {2015})}\BibitemShut
  {NoStop}%
\bibitem [{\citenamefont {Kuznetsov}\ and\ \citenamefont
  {Turitsyn}(1988)}]{Kuznetsov1988}%
  \BibitemOpen
  \bibfield  {author} {\bibinfo {author} {\bibfnamefont {E.~A.}\ \bibnamefont
  {Kuznetsov}}\ and\ \bibinfo {author} {\bibfnamefont {S.~K.}\ \bibnamefont
  {Turitsyn}},\ }\href@noop {} {\bibfield  {journal} {\bibinfo  {journal} {Sov.
  Phys. JEPT}\ }\textbf {\bibinfo {volume} {67}},\ \bibinfo {pages} {1583}
  (\bibinfo {year} {1988})}\BibitemShut {NoStop}%
\bibitem [{\citenamefont {Kuznetsov}\ and\ \citenamefont
  {Rasmussen}(1995)}]{Kuznetsov1995}%
  \BibitemOpen
  \bibfield  {author} {\bibinfo {author} {\bibfnamefont {E.~A.}\ \bibnamefont
  {Kuznetsov}}\ and\ \bibinfo {author} {\bibfnamefont {J.~J.}\ \bibnamefont
  {Rasmussen}},\ }\href@noop {} {\bibfield  {journal} {\bibinfo  {journal}
  {Phys. Rev. E}\ }\textbf {\bibinfo {volume} {51}},\ \bibinfo {pages} {4479}
  (\bibinfo {year} {1995})}\BibitemShut {NoStop}%
\bibitem [{\citenamefont {Huang}\ \emph {et~al.}(2003)\citenamefont {Huang},
  \citenamefont {Makarov},\ and\ \citenamefont {Velarde}}]{Huang2003}%
  \BibitemOpen
  \bibfield  {author} {\bibinfo {author} {\bibfnamefont {G.}~\bibnamefont
  {Huang}}, \bibinfo {author} {\bibfnamefont {V.~A.}\ \bibnamefont {Makarov}},
  \ and\ \bibinfo {author} {\bibfnamefont {M.~G.}\ \bibnamefont {Velarde}},\
  }\href@noop {} {\bibfield  {journal} {\bibinfo  {journal} {Phys. Rev. A}\
  }\textbf {\bibinfo {volume} {67}},\ \bibinfo {pages} {023604} (\bibinfo
  {year} {2003})}\BibitemShut {NoStop}%
\bibitem [{\citenamefont {Ku}\ \emph {et~al.}(2014)\citenamefont {Ku},
  \citenamefont {Ji}, \citenamefont {Mukherjee}, \citenamefont
  {Guardado-Sanchez}, \citenamefont {Cheuk}, \citenamefont {Yefsah},\ and\
  \citenamefont {Zwierlein}}]{Ku2014}%
  \BibitemOpen
  \bibfield  {author} {\bibinfo {author} {\bibfnamefont {M.~J.}\ \bibnamefont
  {Ku}}, \bibinfo {author} {\bibfnamefont {W.}~\bibnamefont {Ji}}, \bibinfo
  {author} {\bibfnamefont {B.}~\bibnamefont {Mukherjee}}, \bibinfo {author}
  {\bibfnamefont {E.}~\bibnamefont {Guardado-Sanchez}}, \bibinfo {author}
  {\bibfnamefont {L.~W.}\ \bibnamefont {Cheuk}}, \bibinfo {author}
  {\bibfnamefont {T.}~\bibnamefont {Yefsah}}, \ and\ \bibinfo {author}
  {\bibfnamefont {M.~W.}\ \bibnamefont {Zwierlein}},\ }\href@noop {} {\bibfield
   {journal} {\bibinfo  {journal} {Phys. Rev. Lett.}\ }\textbf {\bibinfo
  {volume} {113}},\ \bibinfo {pages} {065301} (\bibinfo {year}
  {2014})}\BibitemShut {NoStop}%
\bibitem [{\citenamefont {Ku}\ \emph {et~al.}(2015)\citenamefont {Ku},
  \citenamefont {Mukherjee}, \citenamefont {Yefsah},\ and\ \citenamefont
  {Zwierlein}}]{Ku2015}%
  \BibitemOpen
  \bibfield  {author} {\bibinfo {author} {\bibfnamefont {M.~J.}\ \bibnamefont
  {Ku}}, \bibinfo {author} {\bibfnamefont {B.}~\bibnamefont {Mukherjee}},
  \bibinfo {author} {\bibfnamefont {T.}~\bibnamefont {Yefsah}}, \ and\ \bibinfo
  {author} {\bibfnamefont {M.~W.}\ \bibnamefont {Zwierlein}},\ }\href@noop {}
  {\bibfield  {journal} {\bibinfo  {journal} {arXiv:1507.01047}\ } (\bibinfo
  {year} {2015})}\BibitemShut {NoStop}%
\bibitem [{\citenamefont {Freilich}\ \emph {et~al.}(2010)\citenamefont
  {Freilich}, \citenamefont {Bianchi}, \citenamefont {Kaufman}, \citenamefont
  {Langin},\ and\ \citenamefont {Hall}}]{Freilich2010}%
  \BibitemOpen
  \bibfield  {author} {\bibinfo {author} {\bibfnamefont {D.~V.}\ \bibnamefont
  {Freilich}}, \bibinfo {author} {\bibfnamefont {D.~M.}\ \bibnamefont
  {Bianchi}}, \bibinfo {author} {\bibfnamefont {A.~M.}\ \bibnamefont
  {Kaufman}}, \bibinfo {author} {\bibfnamefont {T.~K.}\ \bibnamefont {Langin}},
  \ and\ \bibinfo {author} {\bibfnamefont {D.~S.}\ \bibnamefont {Hall}},\
  }\href@noop {} {\bibfield  {journal} {\bibinfo  {journal} {Science}\ }\textbf
  {\bibinfo {volume} {329}},\ \bibinfo {pages} {1182} (\bibinfo {year}
  {2010})}\BibitemShut {NoStop}%
\bibitem [{\citenamefont {Navarro}\ \emph {et~al.}(2013)\citenamefont
  {Navarro}, \citenamefont {Carretero-{Gonz\'alez}}, \citenamefont {Torres},
  \citenamefont {Kevrekidis}, \citenamefont {Frantzeskakis}, \citenamefont
  {Ray}, \citenamefont {Altuntas},\ and\ \citenamefont {Hall}}]{Navarro2013}%
  \BibitemOpen
  \bibfield  {author} {\bibinfo {author} {\bibfnamefont {R.}~\bibnamefont
  {Navarro}}, \bibinfo {author} {\bibfnamefont {R.}~\bibnamefont
  {Carretero-{Gonz\'alez}}}, \bibinfo {author} {\bibfnamefont {P.~J.}\
  \bibnamefont {Torres}}, \bibinfo {author} {\bibfnamefont {P.~G.}\
  \bibnamefont {Kevrekidis}}, \bibinfo {author} {\bibfnamefont {D.~J.}\
  \bibnamefont {Frantzeskakis}}, \bibinfo {author} {\bibfnamefont {M.~W.}\
  \bibnamefont {Ray}}, \bibinfo {author} {\bibfnamefont {E.}~\bibnamefont
  {Altuntas}}, \ and\ \bibinfo {author} {\bibfnamefont {D.~S.}\ \bibnamefont
  {Hall}},\ }\href@noop {} {\bibfield  {journal} {\bibinfo  {journal} {Phys.
  Rev. Lett.}\ }\textbf {\bibinfo {volume} {110}},\ \bibinfo {pages} {225301}
  (\bibinfo {year} {2013})}\BibitemShut {NoStop}%
\end{thebibliography}%

\end{document}